\documentclass[pra,showpacs,superscriptaddress,amssymb,twocolumn,longbibliography]{revtex4-1}

\usepackage{float}
\usepackage{amsmath,amssymb,setspace}
\usepackage{graphicx}
\usepackage{subfigure}  
\usepackage[usenames]{color}
\usepackage{graphicx}
\usepackage{bm, bbm}      
\usepackage{color}
\usepackage{srcltx}
\usepackage{hyperref}
\hypersetup{
    colorlinks=true,       
    linkcolor=cyan,          
    citecolor=magenta,        
    filecolor=magenta,      
    urlcolor=cyan,           
    runcolor=cyan
}
\usepackage{upgreek}

\usepackage{slashed}

\def\pd{\partial}

\def\sqr#1#2{{\vcenter{\vbox{\hrule height.#2pt \hbox{\vrule width.#2pt height#1pt \kern#1pt \vrule width.#2pt}\hrule height.#2pt}}}}

\def\beq#1{\begin{equation} \label{#1}}
\def\eeq{\end{equation}}
\def\ben{\begin{equation*}}
\def\een{\end{equation*}}
\def\bequa{\begin{eqnarray}}
\def\eequa{\end{eqnarray}}

\def\Tr{\mathop{\mathrm{Tr}}}

\def\b0d{b_0^{\dagger}}

\def\ml{m_{\text{large}}}
\def\ms{m_{\text{small}}}

\newcommand{\braket}[2]{\langle #1 | #2\rangle}
\newcommand{\ket}[1]{ | #1\rangle}
\newcommand{\bra}[1]{\langle #1 | }
\newcommand{\ketbra}[2]{|#1\rangle\langle #2|}

\newcommand{\ignore}[1]{}

\newcommand\eea{\end{eqnarray}}
\newcommand\bea{\begin{eqnarray}}
\newcommand{\bes}{\begin{subequations}}
\newcommand{\ees}{\end{subequations}}

\def\beq{\begin{equation}}
\def\eeq{\end{equation}}

\def\al{\alpha}

\newcommand{\be}{\begin{equation}}
\newcommand{\ee}{\end{equation}}
\newcommand{\ba}{\begin{align}}
\newcommand{\ea}{\end{align}}
\newcommand{\bg}{\begin{gather}}
\newcommand{\eg}{\end{gather}}
\newcommand{\bseq}{\begin{subequations}}
\newcommand{\eseq}{\end{subequations}}

\renewcommand{\ln}{\mathop{\rm ln}\nolimits}

\newcommand{\red}[1]{\textcolor{red}{#1}}
\newcommand{\green}[1]{\textcolor{green}{#1}}

\begin{document}

\title{Quantum annealing correction at finite temperature: ferromagnetic $p$-spin models}
 
\author{Shunji Matsuura}
\affiliation{Niels Bohr International Academy and Center for Quantum Devices,
Niels Bohr Institute, Copenhagen University, Blegdamsvej 17, Copenhagen, Denmark}
\author{Hidetoshi Nishimori}
\affiliation{Department of Physics, Tokyo Institute of Technology, Oh-okayama, Meguro-ku, Tokyo 152-8551, Japan}
\author{Walter Vinci}
\affiliation{Department of Electrical Engineering, University of Southern California, Los Angeles, California 90089, USA}
\affiliation{Department of Physics and Astronomy, University of Southern California, Los Angeles, California 90089, USA}
\affiliation{Center for Quantum Information Science \& Technology, University of Southern California, Los Angeles, California 90089, USA}
\author{Tameem Albash}
\affiliation{Department of Physics and Astronomy, University of Southern California, Los Angeles, California 90089, USA}
\affiliation{Center for Quantum Information Science \& Technology, University of Southern California, Los Angeles, California 90089, USA}
\affiliation{Information Sciences Institute, University of Southern California, Marina del Rey, CA 90292}
\author{Daniel A. Lidar}
\affiliation{Department of Electrical Engineering, University of Southern California, Los Angeles, California 90089, USA}
\affiliation{Department of Physics and Astronomy, University of Southern California, Los Angeles, California 90089, USA}
\affiliation{Center for Quantum Information Science \& Technology, University of Southern California, Los Angeles, California 90089, USA}
\affiliation{Department of Chemistry, University of Southern California, Los Angeles, California 90089, USA}

\date{\today}
\begin{abstract}

The performance of open-system quantum annealing is adversely affected by thermal excitations out of the ground state. While the presence of energy gaps between the ground and excited states suppresses such excitations, error correction techniques are required to ensure full scalability of quantum annealing. Quantum annealing correction (QAC) is a method that aims to improve the performance of quantum annealers when control over only the problem (final) Hamiltonian is possible, along with decoding. 
Building on our earlier work [S. Matsuura \textit{et al.}, Phys. Rev. Lett. \textbf{116}, 220501 (2016)], we study QAC using analytical tools of statistical physics by considering the effects of temperature and a transverse field on the penalty qubits in the ferromagnetic $p$-body infinite-range transverse-field Ising model.  
We analyze the effect of QAC on second ($p=2$) and first ($p\geq 3$) order phase transitions, and construct the phase diagram as a function of temperature and penalty strength. Our analysis reveals that for sufficiently low temperatures and in the absence of a transverse field on the penalty qubit, QAC breaks up a single, large free energy barrier into multiple smaller ones. We find theoretical evidence for an optimal penalty strength in the case of a transverse field on the penalty qubit, a feature observed in QAC experiments. 
Our results provide further compelling evidence that QAC provides an advantage over unencoded quantum annealing.  
\end{abstract} 
\maketitle


\section{Introduction}

Quantum annealing (QA) is a heuristic optimization method that minimizes classical cost functions using quantum adiabatic evolutions \cite{finnila_quantum_1994,kadowaki_quantum_1998,farhi_quantum_2001,brooke_tunable_2001,Santoro,RevModPhys.80.1061}. For closed systems, adiabaticity guarantees that the time-evolved state is an instantaneous ground state of the time-dependent QA Hamiltonian \cite{Kato:50,Jansen:07,lidar:102106}, 
which in turn guarantees that the final state is a solution of the optimization problem. A non-adiabatic evolution causes transitions into excited states, which  correspond to computational errors. In open systems, coupling to the environment causes further errors even if the evolution is perfectly adiabatic. 
Coupling to a thermal environment results in excitations at any finite temperature \cite{childs_robustness_2001,PhysRevLett.95.250503,TAQC,amin_decoherence_2009,ABLZ:12-SI,Albash:2015nx}. 

While QA is believed to be robust against certain types of decoherence \cite{childs_robustness_2001,PhysRevLett.95.250503,TAQC,amin_decoherence_2009,ABLZ:12-SI,Albash:2015nx}, QA remains vulnerable to the aforementioned thermal excitations that depopulate the ground state, so quantum error correction is necessary for scalability.
Currently, despite considerable theoretical progress in the development of quantum error suppression and correction for QA \cite{jordan2006error,PhysRevLett.100.160506,PhysRevA.86.042333,Young:13,Sarovar:2013kx,Ganti:13,Young:2013fk,Bookatz:2014uq,Marvian:2014nr,Jiang:2015kx,Marvian:2016kb,Marvian-Lidar:16}, an adiabatic version of the accuracy-threshold theorem (see, e.g., Ref.~\cite{Aliferis:05}) has not yet been established.
While it is clearly important to address the theoretical fault tolerance challenge \cite{Mizel:2014sp}, there has been a great deal of interest in investigating implementable error correction methods on near-term devices.  One motivation for this is the commercial availability of quantum annealing hardware, the D-Wave processors \cite{Dwave,Harris:2010kx,Bunyk:2014hb}, which are known to be prone to precision and thermal errors \cite{q108,SSSV,Albash:2014if,Crowley:2014qp,Vinci:2014fk,Martin-Mayor:2015dq}. Error correction methods for QA, known as quantum annealing correction (QAC), have been developed and demonstrated on the D-Wave processors \cite{PAL:13,PAL:14,Vinci:2015jt,Mishra:2015,vinci2015nested}. 

In QAC, the problem Hamiltonian is encoded using a quantum error detection code. Excitations out of the ground state are suppressed via the introduction of energy penalties that commute with the encoded problem Hamiltonian. 
A classical post-processing (decoding) step allows for further recovery of the logical ground state even after such excitations have occurred. We distinguish between two versions of QAC. In one case designated penalty qubits are used to increase the energy of erroneous states \cite{PAL:13,PAL:14,Vinci:2015jt,Mishra:2015}, while in the other case there are no designated penalty qubits but some Ising couplings are used to impose energy penalties \cite{Vinci:2015jt,Mishra:2015,vinci2015nested}.  Here we focus on the former and reserve a study of the latter for a future publication \cite{Matsu2016-part2}.
 
In a previous paper \cite{MNAL:15}, we investigated how QAC affects the success rate of QA using mean field methods. Focusing mostly on the zero temperature case, we showed that QAC can remove or weaken phase transitions during the 
the evolution by increasing the energy gap. As a consequence the success rate for QA is expected to improve when using QAC.  Indeed, this is what was observed experimentally \cite{PAL:13,PAL:14,Vinci:2015jt,Mishra:2015}. 

The main goal of this paper is to extend the analysis of QAC to the finite temperature case.
To accomplish this we use analytical methods borrowed from statistical physics, namely 
Landau-Ginzburg theory applied to the free energy derived for the quantum model after 
the Suzuki-Trotter procedure. In this setting we analyze the phase transitions associated with QAC at finite temperature, since the nature of the phase transition often determines computational complexity: typically first order transitions are associated with exponentially small gaps, while second order transitions feature polynomially small gaps (some exceptions wherein a first order quantum phase transition is associated with a polynomially small gap are known \cite{CabreraJullien-Gap1987,Tsuda:2013br}). We also consider the free energy and study barrier height and width, as these determine the tunneling rate and (as we show) can also be related to the quantum gap; thus they set the inverse time scales for  adiabatic evolution. 

In addition to studying QAC at finite temperatures, a novel aspect of the analysis we present here is a study of the effect of a transverse field acting on the penalty qubits. The presence of the transverse field on the penalty qubits is a feature of all the experiments performed so far of QAC \cite{PAL:13,PAL:14,Vinci:2015jt,Mishra:2015}, but it was ignored for simplicity in our earlier study \cite{MNAL:15}.  
We provide a theoretical justification for the existence of an optimal penalty strength that maximizes the undecoded ground state probability, a feature that was observed in the experiments \cite{PAL:13}.

This paper is organized as follows. In Section~\ref{penalty QAC section} we briefly review QA and QAC.
In Section~\ref{No transverse field in the penalty} we study the free energy and finite-temperature phase transitions without a transverse field on the penalty qubits. We show that the penalty term in general weakens the phase transitions or splits one ``hard" first-order phase transition into multiple, weaker phase transitions. This corresponds to larger energy gaps, which are expected to improve the performance of  QA. Thermal effects, however, have a competing effect, increasing the potential barriers and thus reducing the gaps.  In Section~\ref{transverse field in the penalty}, we consider phase transition with a transverse field on the penalty qubits. We find that this transverse field also increases the potential barrier at the phase transitions. We therefore conclude that adding a transverse field on the penalty qubits reduces the effectiveness of QAC. 
We conclude in Section~\ref{conc}. Additional technical details are presented in the Appendix.

\section{Quantum Annealing and Quantum Annealing Correction}
\label{penalty QAC section}

Quantum annealing is designed to solve combinatorial optimization problems. It is formulated in terms of a classical Ising Hamiltonian of the form
\bea
H^{Z}=-\sum_{i}h_{i}\sigma^{z}_{i}-\sum_{(i,j)}J_{ij}\sigma^{z}_{i}\sigma^{z}_{j}\ ,
\label{problem Hamil}
\eea
where $i\in [1,N]$ are the spin sites, $\{h_{i}\}$ are the local fields, $\{ J_{ij} \}$ are the Ising couplings, and $\sigma^{z}_{i}$
is the $z$ component of the Pauli matrices acting on site $i$.
The optimization problem is encoded into the parameters $h_{i}$ and $J_{ij}$.
The search space is the set of classical spin configurations, and the optimal solution is the ground state or one of the ground states of the Ising Hamiltonian.
For general $h_i$ and $J_{ij}$, finding a ground state of the Hamiltonian above is NP-hard \cite{Barahona1982}. In physical terms, the Hamiltonian encodes a ``rough" energy landscape so that heuristic search methods such as simulated annealing \cite{kirkpatrick_optimization_1983} tend to become trapped in local minima. QA uses quantum fluctuations to find a ground state of the ``problem Hamiltonian" $H^Z$. Quantum fluctuations are induced by a simple ``driver Hamiltonian" $H^X$ whose ground state is easy to prepare since it involves no interactions. We consider the standard transverse-field driver Hamiltonian
\bea
H^{X}=-\sum_{i=1}^N\sigma^{x}_{i}\ .
\label{driver Hamil}
\eea
The time-dependent QA Hamiltonian is given by
\bea
H(t)=A(t)H^{X}+B(t)H^{Z},~~~~t\in[0,t_f]\ .
\label{driver functs}
\eea
The driver term dominates the Hamiltonian at the initial time $t=0$: $A(0)\gg B(0)$.
Then $A(t)$ decreases while $B(t)$ increases, and at the end of the 
evolution
$t=t_f$, the problem Hamiltonian dominates: $A(t_f)\ll B(t_f)$. In the absence of any coupling to an external environment and provided the evolution is slow (adiabatic) compared to the timescale set by the minimum inverse gap of $H(t)$, the final state is a ground state of $H(t_f)$ with high probability and gives the solution to the optimization problem.

In the open system case the steady state is no longer the ground state, and the spectral gap of the Liouvillian sets the inverse timescale for adiabatic evolution \cite{joye_general_2007,oreshkov_adiabatic_2010,Avron:2012tv,Venuti:2015kq}. 
Coupling to the environment (typically modeled as a thermal bath) introduces computational errors in the form of thermally induced excitations \cite{childs_robustness_2001,PhysRevLett.95.250503,TAQC,amin_decoherence_2009,ABLZ:12-SI,Albash:2015nx}. Error correction is thus necessary in any physical implementation of QA. In the QAC approach,  the quantum state is protected in three steps.
First, a classical repetition code is used, whereby a logical qubit is encoded into $C$ physical qubits. The problem Hamiltonian is correspondingly ``encoded", e.g., every Pauli operator is replaced by the corresponding encoded Pauli operator. 
In the second step we deform the Hamiltonian by adding penalty terms that (i) commute with the problem Hamiltonian and (ii) anticommute with bit-flip errors. Step (i) ensures that the encoded problem Hamiltonian and the penalty terms share the same set of eigenstates, and step (ii) ensures that it costs more energy for the environment to generate any errors that anti-commute with $\sigma^z$.
In the third step the logical qubits are decoded at the end of the 
evolution, either by majority vote \cite{PAL:13,PAL:14,Vinci:2015jt,Mishra:2015} or energy minimization \cite{Vinci:2015jt,Mishra:2015}.

The encoded Hamiltonian for QAC that we consider is of the form
\bea \label{eqt:QACHamiltonian}
H=\sum_{c=1}^{C}(H^{Z}_{c}+\Gamma H_{c}^{X}+\gamma H^{P}_{c})\ ,
\eea
where $H^{Z,X}_{c}$ are $C$ identical copies of the same $H^{Z,X}$  defined in Eqs.~\eqref{problem Hamil} and \eqref{driver Hamil}.
The third term $H^{P}_{c}$ is the penalty term, with $\gamma$ its strength. $\Gamma$ is the strength of the quantum term which sweeps from $\infty$ to zero
during the quantum annealing evolution. 

As in our earlier work \cite{MNAL:15}, we choose the penalty as the sum over the stabilizer elements of the  repetition code, anticommuting with all single-qubit bit-flip errors, as required for error detection and suppression of such errors  \cite{Gottesman:1996fk}:
\bea 
H^{P}_{c}=-\sum_{i=1}^N\sigma_{ic}^{z}\sigma_{i0}^{z}\ .
\eea
The penalty term includes an independent penalty qubit $\sigma_{i0}^{z}$ for each logical qubit $i$.
It couples ferromagnetically to all $C$ physical qubits corresponding to logical qubit $i$, thus creating an energy penalty for misalignment with the penalty qubit. 
Note that the degenerate ground states of the penalty term are $|\bar{0}  \rangle_{i} |0\rangle_{P}$ 
and $|\bar{1}  \rangle_{i} |1\rangle_{P}$ where $|0\rangle_{P}$ and $|1\rangle_{P}$ are the states of the penalty qubit, and the code space of the $i^\textrm{th}$ logical qubit is spanned by the two states with all the ``problem spins" pointing in the same direction: $\ket{\bar{0} }_i=\ket{00 \cdots 0}_{i} $ and $\ket{\bar{1} }_i=\ket{11 \cdots 1}_{i}$, where $\ket{0}$ and $\ket{1}$ are the eigenstates of $\sigma^{z}$. 
Therefore, the ground state of the QAC Hamiltonian $H$ at the end of evolution
($\Gamma =0$) is the ground state of
the problem Hamiltonian $H^{Z}$.
However, the ground state of $H^{Z}+\Gamma H^{X}$ is not an eigenstate of $H^{P}=\sum_c H_c^P$.
Thus, although the above form of the penalty term is natural from the problem Hamiltonian point of view, it is not \textit{a priori} clear that it helps the performance of QA.
To ensure an increase in the ground state gap of $H^{Z}+\Gamma H^{X}$
one needs to have a time- and problem-dependent penalty term, or encode the driver Hamiltonian as well \cite{jordan2006error}.

We shall consider a modification of the QAC Hamiltonian in Eq.~\eqref{eqt:QACHamiltonian} by the inclusion of a transverse field on the penalty qubit.  This addition takes the form 
\beq
V=-\epsilon C \Gamma\sum_{i=1}^N\sigma_{i0}^{x}\ ,
\label{eq:V}
\eeq 
where $\epsilon$ is the strength of transverse field on the penalty qubit relative to the problem qubits.  
All QAC experiments to date \cite{PAL:13,PAL:14,Vinci:2015jt,Mishra:2015,Vinci:2015jt,Mishra:2015} were performed in the presence of this term, and one needs to account for this feature in order to have a more complete understanding of the experimental results.

\section{QAC without a transverse field on the penalty qubits}
\label{No transverse field in the penalty}

Let us first consider QAC at finite temperature in the absence of a penalty transverse field, i.e., with $\epsilon=0$ in Eq.~\eqref{eq:V}.
Our problem Hamiltonian is a fully connected $p$-body ferromagnet, i.e.,
\beq
H^{Z}_{c} = -N \left({1\over N}  \sum_{i=1}^{N}\sigma_{ic}^{z}   \right)^{p}\ ,
\eeq
where $i$ is the logical qubit index and $c$ is the copy index. The encoded QAC Hamiltonian~\eqref{eqt:QACHamiltonian} is therefore:
\begin{align}
H&=
-{N}\sum_{c=1}^{C}\left({1\over N}  \sum_{i=1}^{N}\sigma_{ic}^{z}   \right)^{p}
-\Gamma \sum_{c=1}^{C} \sum_{i=1}^{N} \sigma_{ic}^{x}\notag \\
& \quad -\gamma \sum_{c=1}^{C} \sum_{i=1}^{N} \sigma_{ic}^{z}\sigma_{i0}^{z} \ .
\label{hamil-QAC}
\end{align}
It is clear that the ground state of the problem Hamiltonian $H^{Z}_{c}$ is the state with all the spins pointing in the $+z$ direction for odd $p$, and is degenerate with all the spins pointing either in the $+z$ or $-z$ direction for even $p$.
This fact does not change after we turn on the penalty $\gamma$.
The sign of $\gamma$ is not important since the penalty spin $\sigma_{i0}^{z}$ couples only to
the problem qubits $\sigma_{ic}^{z}$, and in the ground state either the penalty and the problem qubits are
parallel ($\gamma>0$) or anti-parallel ($\gamma<0$).
Without loss of generality we choose $\gamma$ to be non-negative.

Phase transitions occurring during the evolution are the main bottlenecks hindering the performance of the algorithm.  Therefore, to obtain useful insights for the mechanisms that help QAC in improving the performance of QA, we analyze the phase transitions associated with the QAC encoding defined in Eq.~\eqref{hamil-QAC}. In particular, we study how such phase transitions are affected by both the presence of a finite temperature and a transverse field on the penalty qubits.

The free energy of this system can be computed using the Suzuki-Trotter decomposition.
In this decomposition, one can transform a quantum problem in $d$ dimensions
into a classical problem in $(d+1)$ dimensions by introducing a Trotter direction \cite{Suzuki:1976rt}.
For our model, one can write the partition function $Z=\Tr e^{-\beta H}$ as a
path integral of an order parameter $m_c={1\over N}\langle \sum_{i} \sigma_{ic}^z\rangle$ along the Trotter direction.  In the large $N$ limit, one can use a static approximation in which $m_c$ does not change along the Trotter direction.  

As shown in Appendix \ref{sec:TransverseFieldExtra}, the saddle point equation for $m_c$ is 
\begin{align}
m_c= 
\frac{\sum_{s\in\{-,+\}} \frac{v_{s,c}}{Q_{s,c}} \sinh(\beta{Q_{s,c}})
\prod_{c' \neq c} \cosh(\beta{Q_{s,c'}}) }
{\prod_{c'=1}^C \cosh\beta{Q_{-,c'}}
+
\prod_{c'=1}^C \cosh\beta{Q_{+,c'}}
} \ .
\label{saddle point eq0}
\end{align}
where
\bes
\bea
\label{eq:Q}
Q_{s,c} &=& \sqrt{v_{s,c}^2+\Gamma^2}\ , \\
v_{s,c}(m) &=& p m_c^{p-1}+s \gamma \ .
\eea
\ees
The minimum free energy solution is given by taking all order parameters equal, $m_c = m, \forall c$, with $m$ satisfying
\begin{align}
m= 
{\sum_{s\in\{-,+\}}\sinh(\beta{Q_s})
\cosh^{C-1}(\beta{Q_s}) {v_{s}\over{Q_s}}
\over
(\cosh\beta{Q_-})^C
+
(\cosh\beta{Q_+})^C
} \ ,
\label{saddle point eq1}
\end{align}
where we dropped the $c$ subscript on $Q_{\pm,c}$ and $v_{\pm,c}$. Considering only this solution, the resultant free energy $F$, defined by $Z=\exp(-\beta N F)$, is
\begin{align}
\label{free energy mean field}
F/C=(p-1)m^p- {1\over C\beta} 
\ln 
\sum_{s\in\{-,+\}} 
\left[ 2\cosh(\beta Q_s) \right]^C\ .
\end{align}
Details of this calculation can be found in the Supplemental Material of Ref.~\cite{MNAL:15} [in particular Eq.~(19) there becomes Eq.~\eqref{free energy mean field}].

\subsection{The zero temperature limit}
Note that when $p$ is even, $Q_\pm$ and hence both $F$ and the RHS of Eq.~\eqref{saddle point eq1} are invariant under the reflections $m \leftrightarrow -m$
and $\gamma \leftrightarrow -\gamma$.
At zero temperature ($\beta\to\infty$), only the largest term in the exponents in Eq.~\eqref{free energy mean field} survives.
By denoting $v^2=\max\{v^2_{+},v^2_{-}\}$, the free energy simplifies to
\bea
F/C=(p-1)m^{p}-\sqrt{v^2+\Gamma^2}\ ,
\label{zero T free energy}
\eea
and the saddle point equation simplifies to 
\bea
m={v \over \sqrt{v^2+\Gamma^2 }}\ .
\eea
For $|m| \approx 0$ we have, using $v= p |m|^{p-1}+ \gamma$, and keeping only the leading order term in $|m|$, 
\bea
\sqrt{v^2+\Gamma^2} &\approx &\sqrt{\gamma^2+\Gamma^2} \left[1+2\gamma p |m|^{p-1}/(\gamma^2+\Gamma^2)\right]^{1/2} \notag \\
&\approx& \sqrt{\gamma^2+\Gamma^2}+\frac{\gamma p |m|^{p-1}}{\sqrt{\gamma^2+\Gamma^2}} \notag \ .
\eea
Therefore, when $|m| \approx 0$:
\beq
\left(F/C \right)_{\beta \to \infty}=-\sqrt{\gamma^2+\Gamma^2}-{p\gamma \over \sqrt{\gamma^2+\Gamma^2}}|m|^{p-1}+\mathcal{O}(m^{p})\ .
\label{eq:F/C-m-small}
\eeq

\subsection{Second order phase transition: $p=2$}

Let us first study the $p=2$ case at finite temperature. 
If there is no penalty term ($\gamma=0$), the phase diagram is simply that of the transverse field Ising model.
At zero temperature [Eq.~\eqref{zero T free energy}], 
the ground state is determined by the dominance of either the ferromagnetic coupling or the quantum fluctuations ($\Gamma$).
If the ferromagnetic coupling dominates (small $\Gamma$), the state is in the symmetry-broken (ferromagnetic) phase ($m\neq0$).
On the other hand, if the quantum fluctuations dominate (large $\Gamma$), the state is in the symmetric (paramagnetic) phase ($m=0$).
These two phases are separated by a second order phase transition at a critical value $\Gamma_c$.

The penalty term induces symmetry breaking. As mentioned above, the ground state configuration for the 
penalty term has both the problem spins and the penalty spin parallel along the $z$ direction.
The symmetric configuration in which the problem spins point in the $x$ direction costs more energy.
In Ref.~\cite{MNAL:15}, it was shown that the symmetry is always broken in the zero temperature limit in the presence of the penalty term.
Indeed, one can easily see this by considering the free energy around $m=0$, given in Eq.~\eqref{eq:F/C-m-small}.  The presence of the linear term shows that for any finite value of $\gamma>0$, the origin $m=0$ is unstable and the ground state is realized at $|m|>0$.
  
\begin{figure}[t]
\includegraphics[scale=0.75]{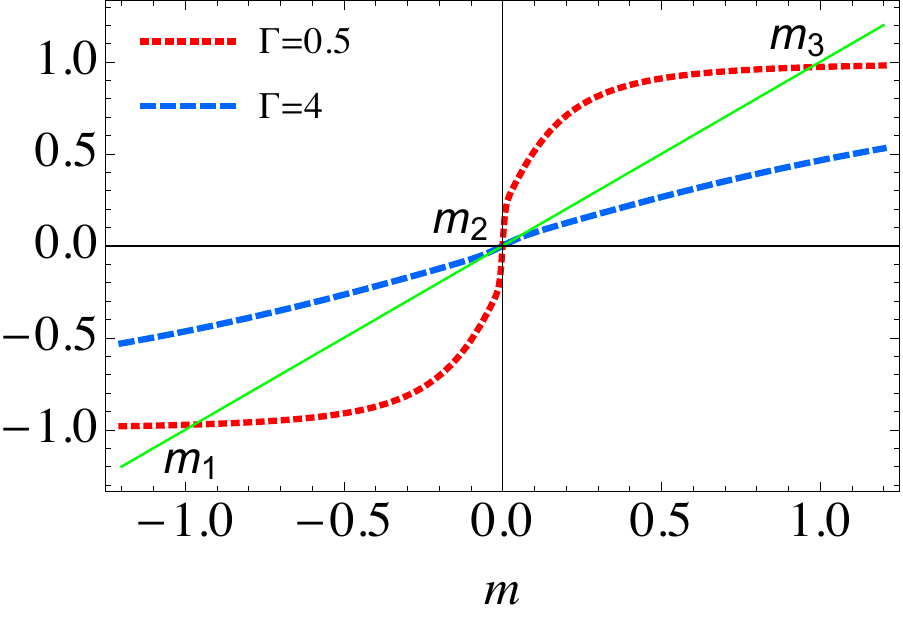}
\caption{Plot of Eq.~\eqref{saddle point eq1} for 
$C=3,p=2,\gamma=0.1,\beta=100$ and $\Gamma=0.5$, $\Gamma=4$. Since $\beta < \infty$,
$m=0$ is always a solution. The number of solutions is determined by the slope of the curves at $m=0$: if the slope is greater than 1, there are three solutions ($m_1 < 0, m_2 = 0, m_3 > 0$), and the state is in the symmetry-broken  phase. If the slope is less than 1, there is only one solution at $m=0$, and the system is in the symmetric phase.}
\label{mM plot}
\end{figure}

At finite temperature, thermal fluctuations can change the situation qualitatively.
From Eq.~\eqref{saddle point eq1},
one can see that $m=0$ is always a solution of the saddle point equation.
The left and the right hand sides of Eq.~\eqref{saddle point eq1} are plotted in
Fig.~\ref{mM plot}.
If the slope of the RHS of Eq.~\eqref{saddle point eq1} at $m=0$ is less than 1, then $m=0$
is the unique solution which suggests that the system is in the symmetric phase with $m=0$.
If the slope is larger than 1, then there are two additional solutions with $m\neq0$. In this case
the system is in the symmetry-broken phase.
At finite temperature, one can see that the critical value of $\Gamma$ is finite.
The Taylor expansion of the right hand side of 
Eq.~\eqref{saddle point eq1} around $m=0$ is
\bea
&\left({2\beta\gamma^2 \over \gamma^2+\Gamma^2}
+{2\Gamma^2 \tanh(\beta\sqrt{\gamma^2+\Gamma^2})\over (\gamma^2+\Gamma^2)^{3/2}}  \right. \nonumber \\
& \left. +{4\beta \gamma^2\tanh^2(\beta\sqrt{\gamma^2+\Gamma^2})\over \gamma^2+\Gamma^2}
\right)m+\mathcal{O}(m^2) \ .
\label{saddle Taylor}
\eea
At sufficiently low temperatures such that $\beta\gamma\gg1$, the coefficient of $m$ is greater than 1 for small $\Gamma$,
while it goes to zero as $\Gamma\to\infty$.
Therefore, the large $\Gamma$ and the small $\Gamma$ regions are separated by a phase transition, and the critical 
value is finite: $0<\Gamma_c<\infty$.

\begin{figure}[h]
      \includegraphics[scale=0.75]{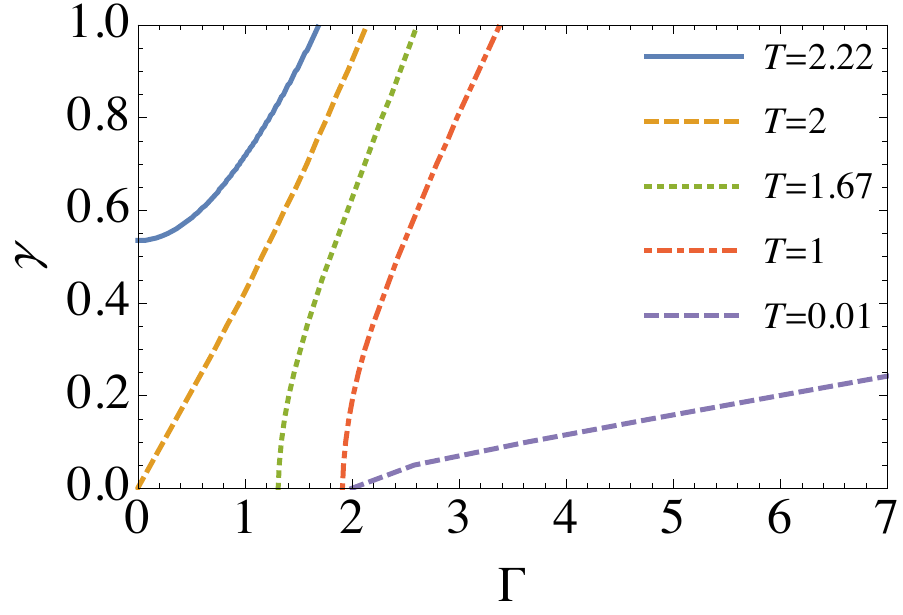}
\caption{Phase diagram $(\Gamma,\gamma)$ at $p=2$ for various values of temperature $T$, and for $C=3$.
Each second order phase transition line separates the symmetry-broken phase $m\neq0$ (above the lines)
from the symmetric phase $m=0$ (below the lines).
For $T< T_1 =1$, the phase transition lines merge at $\Gamma=2$ as $\gamma\to0$.
The $\Gamma(\gamma\to 0)$ point takes smaller values for $T_1 < T < T_2$,
and at $T = T_2=2$ it is zero.
Above $T_2$, the phase transition lines merge on finite $\gamma$ in $\Gamma\to0$ limit.
This means that even in the absence of quantum fluctuations, the spin fluctuations due to the temperature are large enough to have $m=0$ always be the global minimum if $\gamma$ is not too large. 
     }
  \label{gammaGamma p=2} 
\end{figure}

To illustrate these considerations, Fig.~\ref{gammaGamma p=2} shows the phase transition line in the $(\Gamma,\gamma)$ plane for various temperatures.
The region above each fixed temperature line (larger $\gamma$) is the symmetry-broken phase while that below (smaller $\gamma$) is the symmetric phase. The symmetry-broken phase is where the system solves the (trivial) computational problem of finding the ground state with high probability, while the symmetric phase is where it does not. Therefore, from the point of view of successful QAC, we would like the system to end up in the symmetry-broken phase.

There are several interesting and noteworthy features; in particular, we observe the existence of two critical temperatures, $T_1 = 1$ and $T_2 = 2$. For temperatures below $T_1$, the phase transition lines converge to $\Gamma_c=2$ as $\gamma\to 0$.  In the zero temperature limit, the slope of the phase transition line goes to zero, corresponding to the disappearance of the symmetric phase, i.e., an arbitrarily small penalty $\gamma$ suffices to push the system into the symmetry-broken phase.  Conversely, for a fixed $\gamma$, the phase transition happens at larger $\Gamma$ (earlier in the anneal) as the temperature decreases. 

For temperatures between $T_1$ and $T_2$, the $\Gamma_c(\gamma\to 0)$ point moves to values lower than 2. At $T_2$, the phase transition line intersects the origin, i.e., $\Gamma_c(\gamma\to 0)=0$. For temperatures higher than $T_2$, the phase transition lines do not reach $\gamma=0$.  Therefore, there is no phase transition for $\gamma$ below a certain value, and the system is always in the symmetric phase (and fails to solve the computational problem) due to thermal fluctuations.  An example of such a case ($T=2.22$) is shown in Fig.~\ref{gammaGamma p=2}, where there is no phase transition for $\gamma<0.53$.

\subsection{First order phase transitions: $p\ge3$}
\label{sec:1PTp>2}

For $p\ge3$, the phase transition becomes first order, characterized by a discrete jump in the value of $m$ that minimizes the free energy.  

\subsubsection{$T=0$}
Let us first review the zero temperature case  \cite{MNAL:15}. The first order phase transition persists as long as $\gamma$ is less than some critical value $\gamma_c(p)$.
At the critical point, the free energy barrier between different ground states becomes smaller for larger values of $\gamma$.  We interpret this as a softening of the phase transition as the penalty coupling increases. The reason for this, as we show in Appendix~\ref{app:Instanton}, is that the gap $\Delta$ of the QAC Hamiltonian can be estimated as
\beq
\Delta \gtrsim e^{-\Delta m \Delta F N} \ .
\label{eq:gap-F}
\eeq
where $\Delta m$ is the width and $\Delta F$ is the height of the free energy barrier at the phase transition. Thus, a smaller free energy barrier translates into a larger quantum gap $\Delta$.

The Taylor expansion of the free energy around $m=0$, Eq.~\eqref{eq:F/C-m-small}, shows that the symmetric phase $m=0$ is always unstable and will not be realized as the ground state.
The phase transition at finite $\gamma$ is therefore a transition between finite values of $m$, which we denote by $\ms$  and $\ml$ ($\ms < \ml$). Both values are continuous in $\gamma$.
In particular, $\ms$ goes to zero continuously as $\gamma\to0$. As $\gamma$ increases, $\ms$ becomes larger and eventually it merges with $\ml$ at the critical value $\gamma=\gamma_c$. For $\gamma>\gamma_c$, the phase transition disappears \cite{MNAL:15}.

\subsubsection{$T>0$}
At finite temperature, the symmetric point $m=0$ becomes locally stable, as can be seen by computing the Taylor expansion of the free energy [Eq.~\eqref{free energy mean field}] at $m=0$:
\bea
F(m)/C=F(0)/C+(p-1)m^{p}+\mathcal{O}(m^{p+1})\ .
\eea
Therefore, phase transitions can happen not only between $\ms$ and $\ml$ but also between $m=0$ and $m_{\text{small}}$, and between $m=0$ and $m_{\text{large}}$.  Furthermore, unlike the zero temperature case, at finite temperature the point $\Gamma=0$ is always separated from $\Gamma=\infty$ by first order phase transitions, similarly to what happens for $p=2$.  However, the penalty $\gamma$ does soften the phase transition, in that it shrinks the free energy barrier separating degenerate minima of the free energy [recall Eq.~\eqref{eq:gap-F}].

\begin{figure*}[t]
\begin{tabular}{ccc}
\subfigure[]{\includegraphics[scale=0.6]{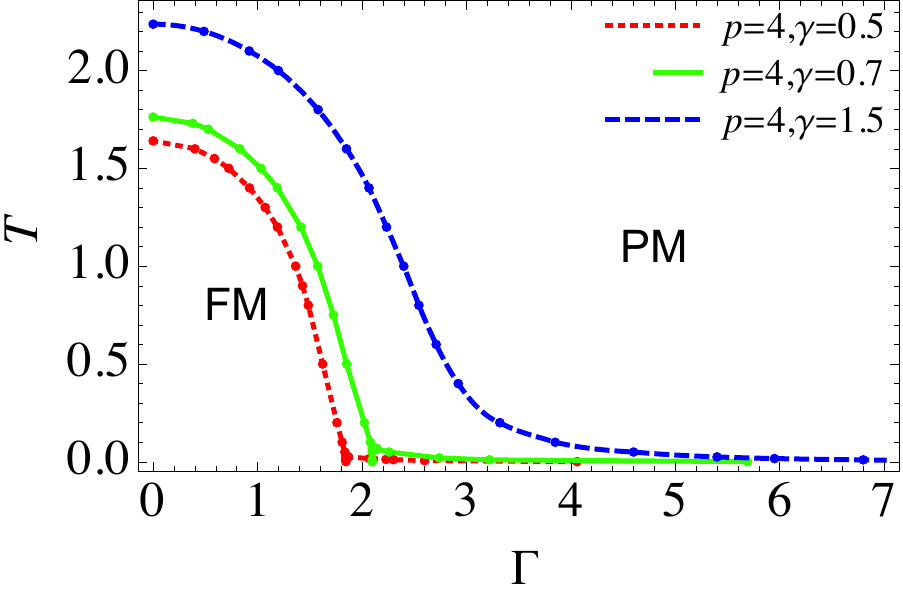} \label{PLpmulRedup4} }&
\subfigure[]{\includegraphics[scale=0.6]{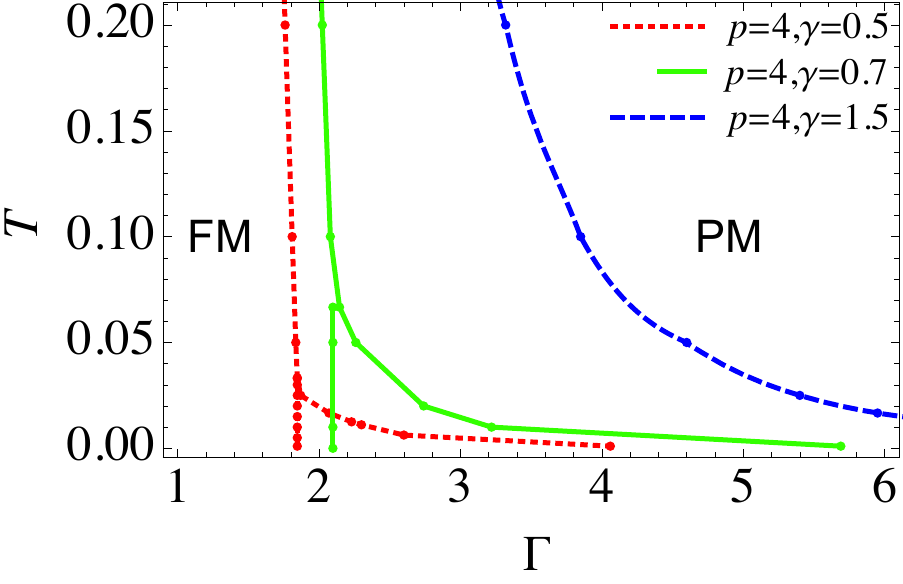} \label{PLpmulRedup4Zoom} }&
\subfigure[]{\includegraphics[scale=0.6]{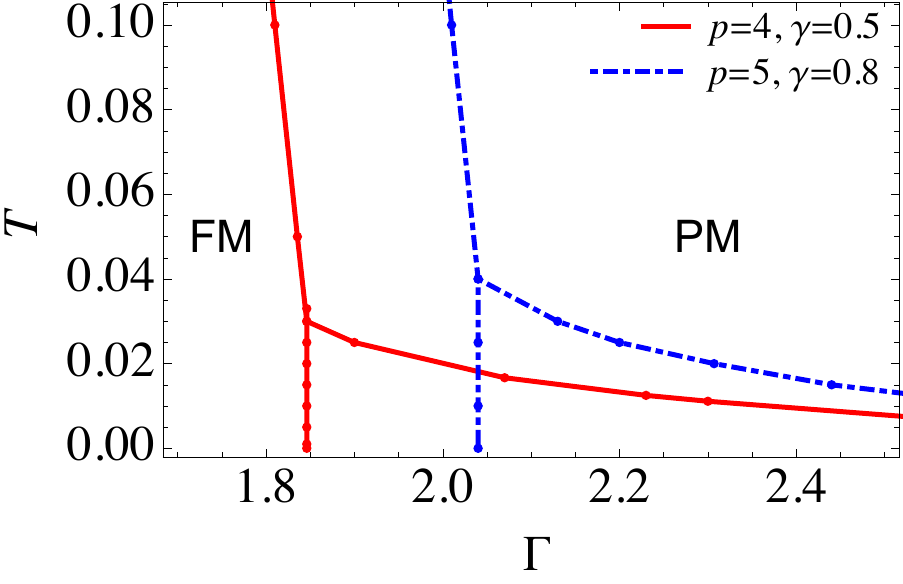}  \label{PLpmulRedu-a} }
\end{tabular}                
\caption{$(\Gamma,T)$ phase diagrams for $p=4,5$ and different penalty values.  The space to the right of the curves corresponds to the paramagnetic or symmetric phase, labeled by `PM', while the space to the left of the curves corresponds to the ferromagnetic or symmetry-broken phase, labeled by `FM'.  (a) For $p=4$ and $\gamma=0.5$, $0.7$  and $1.5$. At zero temperature, there is a first order phase transition. The lower branch ($\Gamma=2.1, T<0.07$) of the phase boundary connects to the phase transition at zero temperature. This line separates the state characterized by $\ml$ from $m_{\text{small}}$.  Another line ($\Gamma>2.1, T<0.07$) is for the phase transition between $\ms$ and $m=0$. There is no corresponding phase transition at $T=0$ and this transition effectively disappears ($\ms$ goes to zero)  as $T$ approaches zero. For $\gamma=1.5$, there is no corresponding phase transition at $T=0$. Therefore there is no branch that reaches $T=0$.  (b) A close-up around the branching point in (a). The branching temperature increases as $\gamma$ increases. (c) The phase diagram for $p=4,\gamma=0.5$ and for $p=5,\gamma=0.8$.}
\label{PLpmulRedu}
\end{figure*}

\begin{figure*}[t]
\begin{tabular}{ccc}
\subfigure[]{     \includegraphics[scale=0.6]{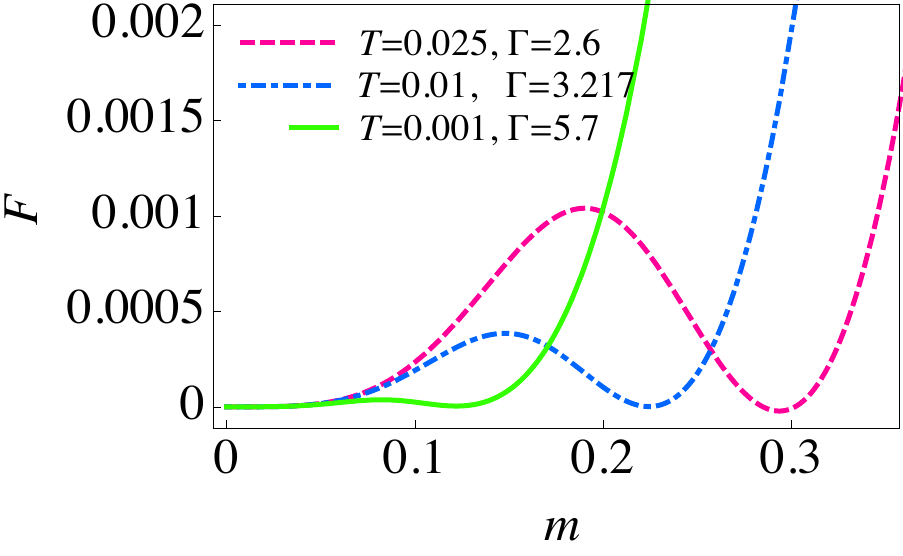} \label{lowT-highG}  }  &
\subfigure[]{     \includegraphics[scale=0.6]{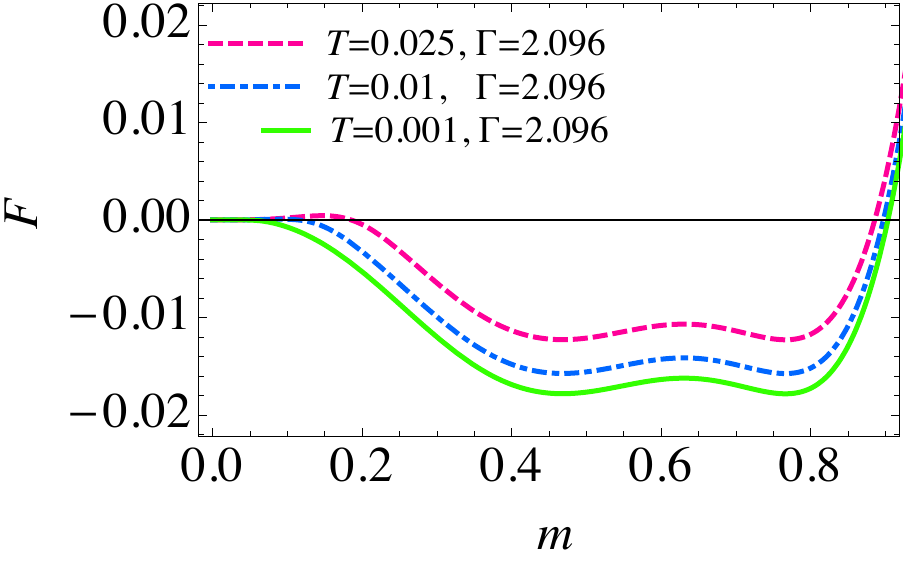}   \label{p4T0025g0a} }     &
\subfigure[]{     \includegraphics[scale=0.6]{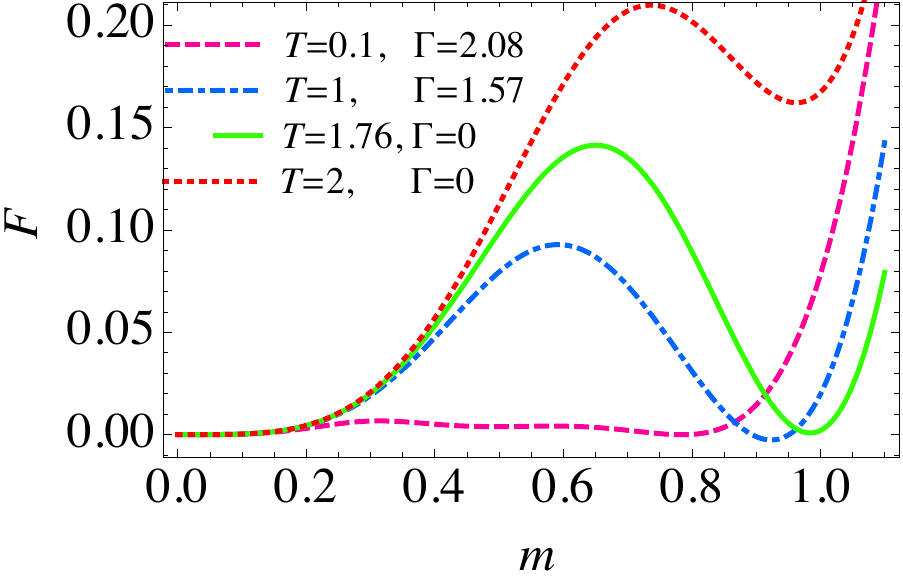}   \label{highTbranch}} \\
\end{tabular}     
\caption{ 
Free energy as a function of the order parameter $m$ along the phase transition lines, for $p=4$,  $\gamma=0.7$. (a) Transition between $m=0$ and $m=m_{\text{small}}$ $(T<T_{\text{branch}},\Gamma>\Gamma_{\text{branch}})$.  (b) Transition between $m=\ms$ and $m=m_{\text{large}}$ $(T<T_{\text{branch}},\Gamma=\Gamma_{\text{branch}})$. (c) Transition between $m=0$ and $m=m_{\text{large}}$ $(T>T_{\text{branch}},\Gamma<\Gamma_{\text{branch}})$.}
  \label{on the transition line}    
\end{figure*}

The $(\Gamma,T)$ phase diagram is shown in Fig.~\ref{PLpmulRedu}.
In Fig.~\ref{PLpmulRedup4}, we plot the phase transition lines for $p=4$ and $\gamma=0.5,0.7,1.5$  ($\gamma_c \simeq 0.8)$.  The point $(\Gamma,T)=(0,0)$, which encodes the solution to the optimization problem, is separated from the large $\Gamma$ and/or the large $T$ region by first order phase transition lines.  When quantum fluctuations are very large $(\Gamma\gg 1)$, the free energy minimum is in the symmetric phase $m=0$.  As quantum fluctuations decrease $(\Gamma\to0)$, the free energy minimum shifts to the ordered phase $m=\ml$ due to the ferromagnetic coupling.  The same thing happens for thermal fluctuations: when the temperature is large, the system is in the symmetric phase $m=0$ and as temperature decreases it crosses into the ordered phase $m\neq0$.  The details of the first order phase transitions are sensitive to the value of the penalty strength $\gamma$, which we we analyze next.
 
\begin{figure}[t]
\includegraphics[scale=0.75]{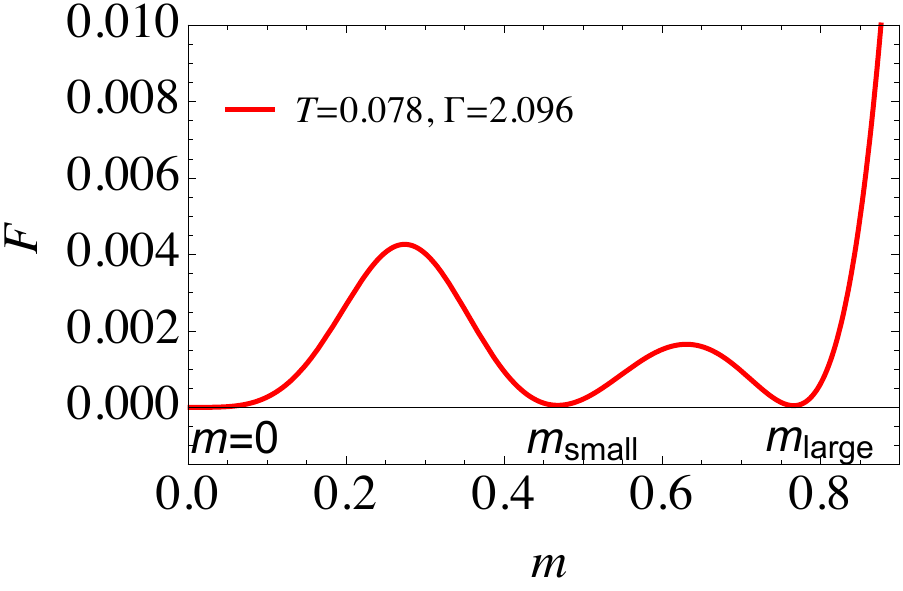} 
\caption{ Free energy at the branching point for $\gamma=0.7$. At $T=0.078$ and $\Gamma=2.096$, the free energy possesses three minima 
at $m=0$, $m=m_{\text{small}}$, and  $m=m_{\text{large}}$. 
 }
\label{branchfreeg07}  
\end{figure}

For $\gamma < \gamma_c$, there is a branch point and three phase transition lines are connected there.  For example, in Fig.~\ref{PLpmulRedup4}, such a branch point can be observed for $\gamma = 0.5$ at $(T,\Gamma) = (0.02, 1.8)$ and for $\gamma = 0.7$ at $(T, \Gamma) = (0.078,2.096)$ [at these points, $F(m=0)$, $F(m_{\text{small}})$, and $F(m_{\text{large}})$ all take the same value at the critical point $\Gamma=\Gamma_c$. See Fig.~\ref{branchfreeg07}]; 
let us denote such ($\gamma$-dependent) branching points by $(T_{\text{branch}},\Gamma_{\text{branch}})$.  The phase transition line $(T<T_{\text{branch}},\Gamma>\Gamma_{\text{branch}})$ indicates the transition between $m=0$ and $m=m_{\text{small}}$. Likewise, 
the phase transition line $(T<T_{\text{branch}},\Gamma=\Gamma_{\text{branch}})$ indicates the transition between $\ms$ and $m=m_{\text{large}}$, and the phase transition line $(T>T_{\text{branch}},\Gamma<\Gamma_{\text{branch}})$ indicates the transition between $m=0$ and $m=m_{\text{large}}$.  This branching behavior is a generic feature for $p\geq 3$, and we show in Fig.~\ref{PLpmulRedu-a} an example for $p = 5$ with $\gamma \leq \gamma_c$.

\subsubsection{Free energies along each phase transition line}

In Fig.~\ref{on the transition line}, we plot the free energies \emph{along} each phase transition line.  
Figure~\ref{lowT-highG} shows the free energy for $(T<T_{\text{branch}},\Gamma >\Gamma_{\text{branch}})$.  As the temperature decreases the critical value of $\Gamma$ becomes larger, and the free energy barrier becomes smaller.  The phase transition effectively disappears at very low temperatures, which is consistent with the fact that there is no corresponding phase
transition at $T=0$.

Figure~\ref{p4T0025g0a}  shows the free energy for $(T<T_{\text{branch}},\Gamma_{\text{branch}}=\Gamma)$.  This transition exists at $T=0$, and the phase transition line at $T>0$ continuously connects to the transition line at $T=0$.  The critical value of $\Gamma$ does not change within numerical precision.

Figure~\ref{highTbranch} shows the free energy for $(T>T_{\text{branch}},\Gamma<\Gamma_{\text{branch}})$.  The critical value of $\Gamma$ becomes smaller as the temperature increases. Above a $\gamma$-dependent critical temperature, the global minimum of the free energy is at $m = 0$, so the state is in the symmetric phase even when there are no quantum fluctuations ($\Gamma=0$).  This state can therefore not recover the solution to the optimization problem.   As can be observed in Fig.~\ref{PLpmulRedup4}, the value of this critical temperature increases with increasing $\gamma$.  We provide additional analysis of the free energy in Appendix \ref{sec:extraF}.

\ignore{
This means that above this temperature the quantum annealing process results in the wrong answer. \red{Can we fix this by increasing $\gamma$? i.e., does a higher $\gamma$ lead to a higher $T_{\text{branch}}$? Or does the $T=0$ result that a penalty larger than $\gamma=0.8$ doesn't help tell us that, at any $T$, increasing $\gamma$ beyond 0.8 doesn't help?}
\green{The value of $T$-intercept in Fig.~\ref{PLpmulRedup4} increases as $\gamma$ becomes large. Namely  the phase transition  exists even at a large value of $T$ as long as $\gamma$ is taken to be large enough. Therefore, the quantum state can be in the correct phase at the end of the evolution ($\Gamma=0$).  The same feature appears in the $p=2$ case (Fig.~\ref{gammaGamma p=2}). 
When $\gamma\ge0.8$ there is only one phase transition. 
yes the $T$-intercept increases as $\gamma$ increases. when $\gamma$ is bigger than 0.8, the branch structure disappears and therefore 
we cannot get a benefit of it. however, we still have a possibility to get the correct answer by increasing $\gamma$ since it prevents the thermal fluctuation, in some sense.}
}

\section{QAC with a transverse field on the penalty qubits}
\label{transverse field in the penalty}

Recall that all the experiments conducted to date on QAC \cite{PAL:13,PAL:14,Vinci:2015jt,Mishra:2015} involve a transverse field on the penalty qubits. In this section we analyze this scenario, defined by the total Hamiltonian that includes the ``penalty transverse field", namely the sum of Eqs.~\eqref{eq:V} and \eqref{hamil-QAC}:
\begin{align}
H =&-{N}\sum_{c=1}^{C}\left({1\over N}  \sum_{i=1}^{N}\sigma^{z}_{ic}   \right)^{p}
-\Gamma\sum_{c=1}^{C}\sum_{i=1}^{N} \sigma^{z}_{ic}
-\epsilon {C} \Gamma \sum_{i=1}^{N} \sigma^{x}_{i0} \notag \\
&-\gamma \sum_{c=1}^{C}\sum_{i=1}^{N} \sigma^{z}_{ic}\sigma^{z}_{i0}\ .
\label{full trans penalty hamil}
\end{align}

Beyond the direct connection to experiments, the addition of the transverse field should make the penalty qubits, and as a consequence the problem qubits, fluctuate more easily, and it is interesting to study whether this helps the QA process cross phase transitions more efficiently.

When a transverse field on the penalty qubits is included, the $C$ copies of the problem qubits and the penalty qubit are no longer effectively decoupled. This means that to understand the effect of the penalty qubits we need to study a $2^{C+1}$-dimensional Hilbert space instead of a $2$-dimensional one.

\subsection{Second order phase transition: $p=2$}

We show in Appendix~\ref{sec:TransverseFieldExtra} that after the Suzuki-Trotter procedure, the free energy at low temperature ($\beta \gg 1$) can be written as
\bea
F/C=(p-1)m^p-{1\over C\beta}\ln\left[ \Tr\exp(\beta H_\mathrm{eff})  \right] \ ,
\label{trans penalty partition func}
\eea
where $H_{\mathrm{eff}}$ is an effective Hamiltonian given by
\beq
H_{\mathrm{eff}} = -(p m^{p}\sigma^{z}_{c}+\gamma
\sigma^{z}_{c}\sigma^{z}_{0}
+ { \Gamma}  \sigma^{x}_{c}+{\epsilon \Gamma } \sigma^{x}_{0}) \ .
\label{eq:Heff}
\eeq
The term $\Tr\exp(\beta H_{\mathrm{eff}})$ can be evaluated by numerically diagonalizing the  $2^{C+1}\times 2^{C+1}$ matrix.  

\begin{figure*}[t]
\subfigure[]{ \includegraphics[scale=0.75]{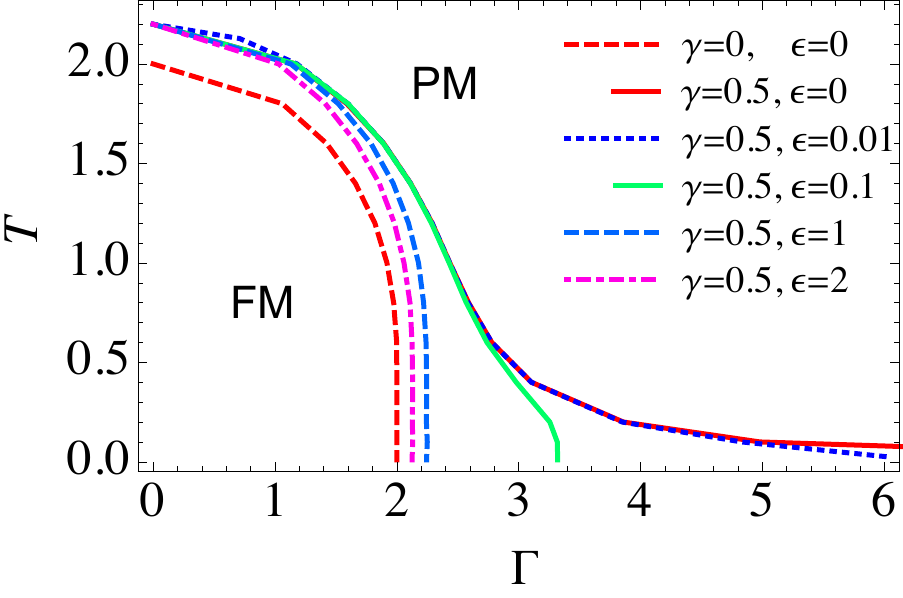}  \label{transp2-a}}\quad \quad \quad \quad \quad \quad
\subfigure[]{ \includegraphics[scale=0.75]{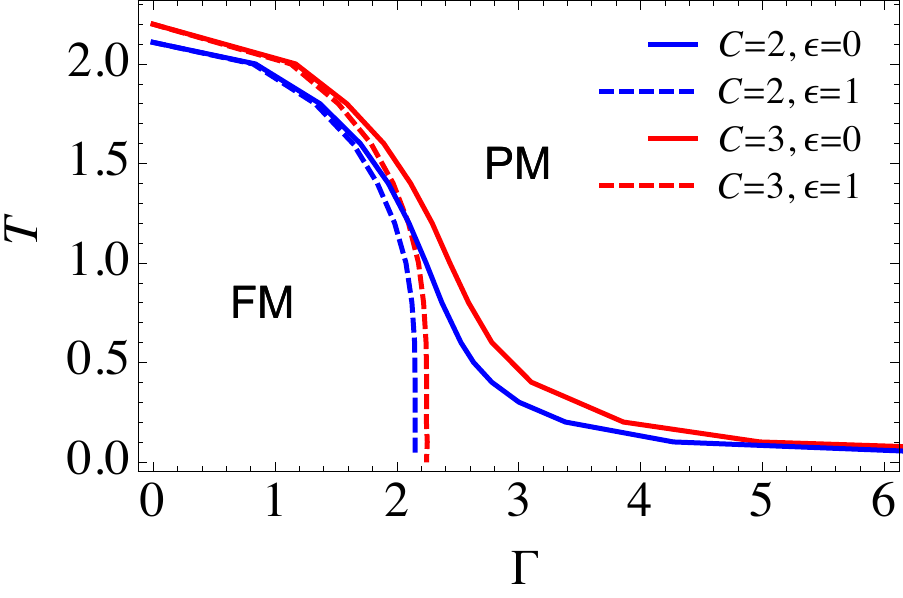} \label{diffCp2g05}}
\caption{(a) Phase diagram in the presence of a penalty transverse field. Shown are the 
$p=2$ phase transition lines with $C=3$ for different values of $\gamma$ and $\epsilon$.
(b) Comparison of $p=2$ phase transition lines with $C=2$ and $C=3$
for $\gamma=0.5$ and $\epsilon=0$ and $1$.  The space to the right of the curves corresponds to the paramagnetic or symmetric phase, while the space to the left of the curves corresponds to the ferromagnetic or symmetry-broken phase.}
 \label{transp2}    
\end{figure*}

The phase diagram in the $(\Gamma,T)$ plane is shown for $p=2$ in Fig.~\ref{transp2-a} for various values of $\gamma$ and $\epsilon$.
The phase transition line separates the symmetry-broken phase $m\neq0$ (below the lines)
from the symmetric phase $m=0$ (above the lines).
As in the case without a penalty transverse field studied in the previous section, there exists a critical temperature above which the state is in the symmetric phase for all values of $\Gamma$ even for finite $\epsilon$. 
However, unlike the $\epsilon = 0$ case, we find that $\Gamma_c$ approaches a finite value in the low temperature limit for finite $\epsilon$.  The phase transition lines at finite $\gamma$ approach that of $\gamma=\epsilon=0$ as $\epsilon$ increases. Thus, the penalty transverse field counterbalances the effect of the penalty term $\gamma$, as it enlarges the phase space of the symmetric phase with $m=0$, where the system fails to solve the optimization problem.

The location of the phase transition depends on the value of $C$. In Fig.~\ref{diffCp2g05}, we compare the phase transition lines for $C=2$ and $3$. While the qualitative features 
for the two $C$ values are the same, the symmetry-broken phase is enlarged as $C$ increases for both the $\epsilon=0$ and $\epsilon=1$ cases. Thus, increasing $C$ acts productively and enhances the phase where the system solves the optimization problem.

\begin{figure}[t]
\includegraphics[scale=0.75]{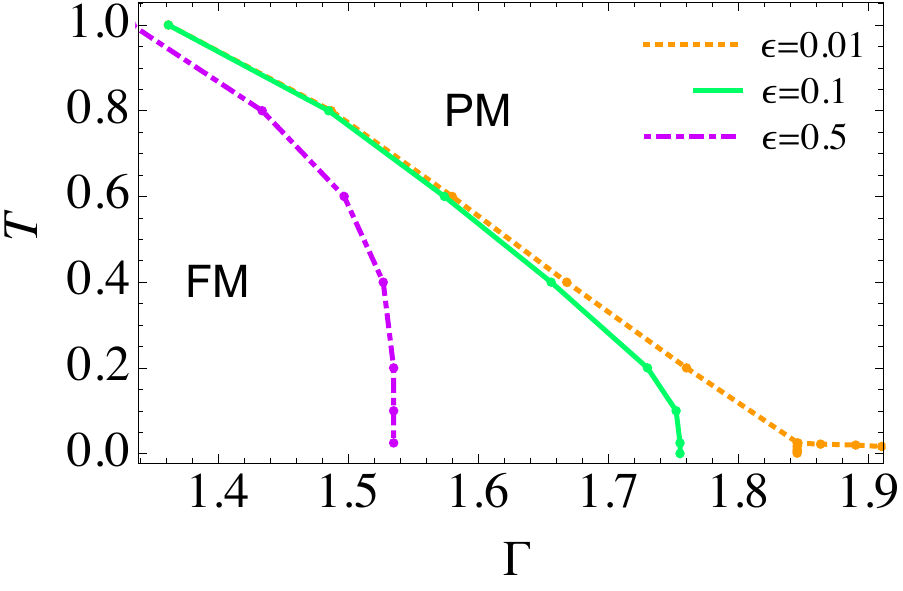}   
\caption{Phase diagram for $p=4$ in the presence of a penalty transverse field.  The space to the right of the curves corresponds to the paramagnetic or symmetric phase, while the space to the left of the curves corresponds to the ferromagnetic or symmetry-broken phase.  The penalty coupling is chosen to be $\gamma=0.5$ for all curves.}
 \label{transg05p4}    
\end{figure}

\begin{figure}[t]
\includegraphics[scale=0.75]{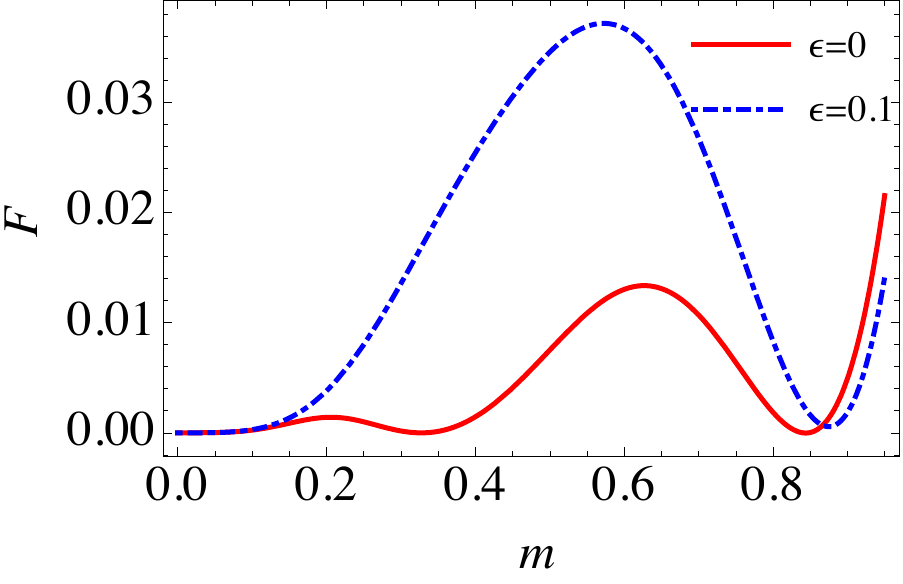} 
\caption{Free energies $F$ with and without the penalty transverse field at the critical points. Parameters are chosen to be $p=4,\gamma=0.5,T=0.03$. The blue line is for  $\epsilon=0,\Gamma=1.85$, and the red line is for $\epsilon=0.1,\Gamma=1.76$. }
\label{compF}    
\end{figure}

\begin{figure*}[t]
\subfigure[]{     \includegraphics[scale=0.6]{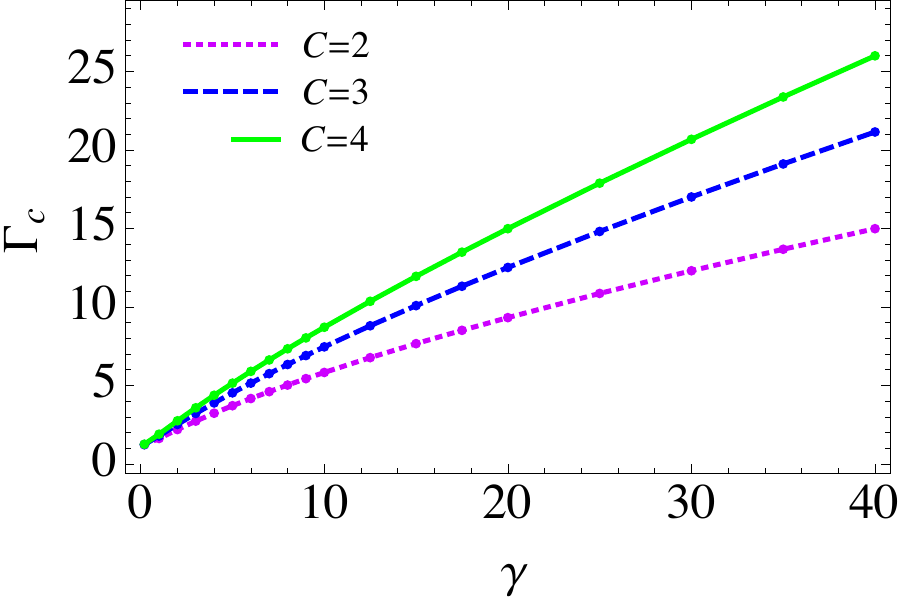}    \label{Plotoptb40e1gG} }
\subfigure[]{     \includegraphics[scale=0.6]{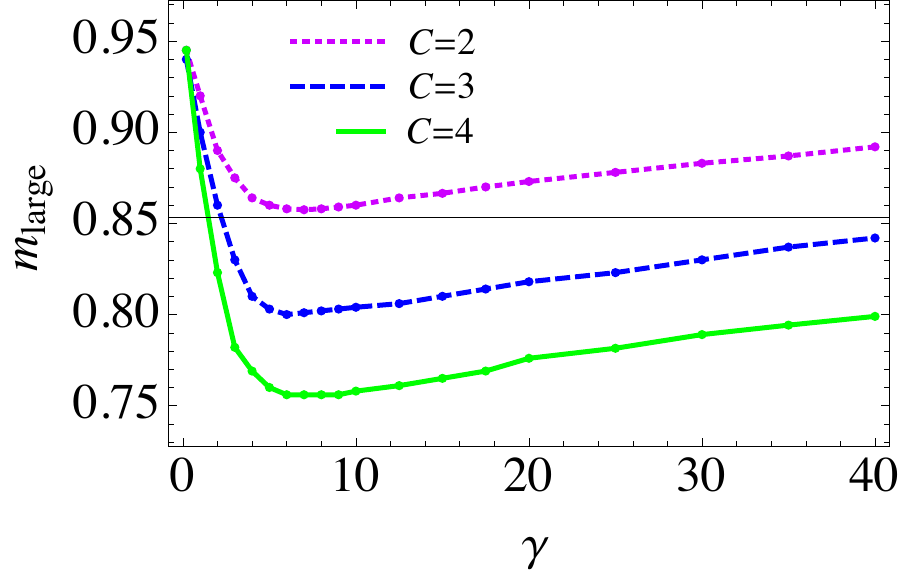}   \label{Plotoptb40e1gm} }
\subfigure[]{     \includegraphics[scale=0.6]{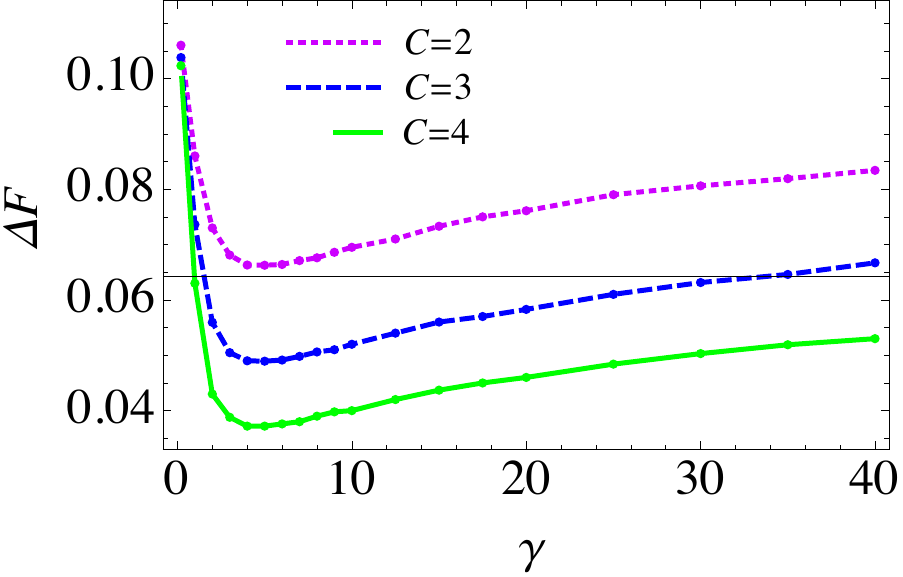}   \label{Plotoptb40e1gF} }
\caption{ (a) Critical value of $\Gamma$ as a function of the penalty coupling $\gamma$.  
(b) $\ml$ as a function of $\gamma$.
(c) the barrier height $\Delta F$  as a function of $\gamma$.
In all panels $p=4$,
$T=0.025$, and $\epsilon=1$.}
\end{figure*}

\subsection{First order phase transitions: $p\ge3$}

The phase diagram in the $(\Gamma,T)$ plane for various values of $\epsilon$ is shown in Fig.~\ref{transg05p4}, for $\gamma=0.5$.
For sufficiently small $\epsilon$, the phase diagram is similar to that of $\epsilon=0$: $\ms$ still exists and can be the lowest free energy state for sufficiently low temperatures. Therefore there are two phase transitions, one from $m=0$ to $\ms$ and the other from $\ms$ to $m_{\text{large}}$.
As the temperature increases, $\ms$ goes to $0$ and above a certain branching point temperature there is only one phase transition from $m=0$ to $m_{\text{large}}$.
At $\epsilon=0.014$ (not shown), this branching point disappears, and there is only one phase transition from $m=0$ to $m_{\text{large}}$.

In order to understand the disappearance of the phase transition as $\epsilon$ is increased, we show a comparison of the free energies for $\epsilon = 0$ and $\epsilon \neq 0$ at their respective  critical $\Gamma$'s in Fig.~\ref{compF}. 
For the choice of parameters shown, there are three local minima for $\epsilon=0$.  As we noted in the previous section, there are correspondingly two phase transitions, one between $m = 0$ and $\ms$ followed by a transition between $\ms$ and $\ml$.
By increasing $\epsilon$, the free energy barrier between $m = 0$ and $\ml$ is raised, and as a consequence the local minimum $\ms$ disappears for sufficiently large $\epsilon$, as shown in Fig.~\ref{compF} for $\epsilon=0.1$.

Recall that as we discussed in Sec.~\ref{sec:1PTp>2}, the splitting of a first order phase transition into multiple smaller phase transitions is beneficial from the QA point of view. 
Further recall that the energy gap between the ground state and the excited state decreases when the potential barrier becomes larger [Eq.~\eqref{eq:gap-F}]. This means that the transverse field decreases the energy gap at the phase transition. Both effects thus adversely affect the success rate of QAC in comparison to having no transverse penalty.

In Appendix~\ref{app:pert-th} we supplement the analysis presented here with perturbation theory in $\epsilon$, which allows us to show analytically how introducing the penalty transverse field removes the local minimum at $\ms$.

\subsection{Optimal value of $\gamma$}

The experimental results reported in Refs.~\cite{PAL:13,PAL:14,Vinci:2015jt,Mishra:2015,vinci2015nested} all included a penalty transverse field and showed a marked improvement in performance relative to the unencoded case. These results required the optimization of the penalty strength in order to maximize the success probability of QAC after decoding, although an optimal penalty strength was also observed when maximizing the physical (i.e., encoded, but undecoded) ground state probability.  Here we do not include a decoding step and instead study the phase transition as a function of $\gamma$ for a fixed value of $\epsilon$ in order to study whether any features of the free energy may indicate the presence of an optimal $\gamma$ value for QAC.

We display the critical value of $\Gamma$, denoted $\Gamma_c$, as a function of $\gamma$ for $\epsilon=1$ in Fig.~\ref{Plotoptb40e1gG}.  $\Gamma_c$ increases monotonically with $\gamma$, so the phase transition happens earlier in the evolution.  (Note that for the value of $\epsilon$ relevant to the experiments, there is only a single phase transition.)
On the other hand, both the free energy height and $\ml$ (which characterizes the free energy barrier width) are non-monotonic functions of $\gamma$ at the critical point, as shown in Figs.~\ref{Plotoptb40e1gm} and \ref{Plotoptb40e1gF}.  The free energy barrier height decreases for $\gamma\le 5$ and increases for $\gamma > 5$.  The barrier width decreases monotonically for $\gamma<6$ but increases monotonically for $\gamma\ge 6$.
Since the barrier height and width control the behavior of the instanton connecting the two free energy minima and hence the tunneling rate at the phase transition point (more details are provided in Appendix \ref{app:Instanton}), the presence of a minimum as a function of $\gamma$ that takes a value smaller than that at $\gamma = 0$ is strong evidence in favor of an optimal $\gamma$ for QAC.
While the position of the minima of the barrier height and width do not coincide precisely, they are broad minima, and it is possible that their combined effect is responsible for the appearance of the optimal $\gamma$ observed for the undecoded ground state in experiments.

\section{Conclusions}
\label{conc}

Building on the methods and results of Ref.~\cite{MNAL:15}, we have extended the mean-field analysis of quantum annealing correction to include the effects of temperature and of a transverse field on the penalty qubits. We find that the effects of a finite temperature $T$ on QA manifests themselves in two different but connected ways, both of which are detrimental. First, the symmetric phase, where quantum annealing fails to find the ground state, is enlarged in the $(\gamma,T)$ phase space, where $\gamma$ is the penalty strength. Second,  the free-energy landscape is modified in such a way that the barrier between the two minima at the phase transition is increased.  This implies that for sufficiently large temperatures the two first-order phase transitions present for the $p>2$ case are replaced by one first-order phase transition with a larger free energy barrier. For first order phase transitions, larger barriers imply smaller tunneling rates between the two minima, which means that the probability of finding the ground state is lowered at fixed annealing time. In fact, the energy gap between the ground state and excited states is a function of the potential barrier, which appears in the exponent for a model with a first order transition [see Eq.~\eqref{eq:gap-F}]. Thus the gap size shrinks exponentially in the free energy barrier height and width.
Clearly, multiple phase transitions with small barriers are preferable to a single transition with a large barrier: in the former case the total annealing time is $t_a^{\text{small}} \sim \max_i 1/\Delta_i  \sim \max_i \exp(\Delta m \Delta F_i^{\text{small}} N)$, while in the latter 
$t_a^{\text{large}} \sim 1/\Delta \sim  \exp(\Delta m \Delta F^{\text{large}}N)$, and $t_a^{\text{small}} \ll t_a^{\text{large}}$ provided $\max_i \Delta F_i^{\text{small}} \ll \Delta F^{\text{large}}$.

We have shown that the introduction of a transverse field on the penalty qubit, for a fixed value of $\gamma$, makes the critical value of the annealing parameter $\Gamma$ smaller for $p=2$, and enlarges the potential barrier for $p\ge2$. In this sense one might say that the penalty transverse field favors the symmetric phase with $m=0$.  

Both the temperature and the additional transverse field favor fluctuations (either of thermal or quantum nature)  that facilitate the appearance of the symmetric phase, so part of our results have an intuitive explanation. On the other hand, intuitively, the transverse field should facilitate the annealing process by inducing quantum fluctuations of the penalty qubit. However, we find that the transverse field tends to reduce the tunneling rate at the phase transitions, a counter-intuitive results. A possible resolution of this puzzle may be that the free energy includes entropy considerations that may not be relevant in the dynamics. Preliminary analysis of dynamics suggest that the transverse field on the penalty qubits accelerates relaxation to equilibrium, possibly by providing quantum transition paths that do not exist in classical dynamics.  It is an important future problem to clarify the balance (or trade-off) between this dynamics and the equilibrium effects associated with the transverse field on the penalty qubit.

Finally, our analysis of the free energy as a function of $\gamma$ revealed a minimum for both the free energy width and barrier height separating minima at the phase transition.  This provides a possible explanation for the observed experimental results reported in Ref.~\cite{PAL:13} of an optimal $\gamma$ for the undecoded (``physical") ground state.  
We postpone the inclusion of decoding to future work \cite{Matsu2016-part2}, which will also include disordered spin models, such as the quantum Hopfield model analyzed for $T=0$ in Ref.~\cite{MNAL:15}.

\begin{acknowledgments}
The work of H.N. was funded by the ImPACT Program of Council for Science, Technology and Innovation, Cabinet Office, Government of Japan, and by the JPSJ KAKENHI Grant No. 26287086. The work of W.V., T.A., and D.L. was supported under ARO Grant No. W911NF- 12-1-0523, ARO MURI Grants No. W911NF-11-1-0268 and No. W911NF-15-1-0582, and NSF Grant No. INSPIRE-1551064.
\end{acknowledgments}

\appendix

\newpage

\section{Free energy for the case of a penalty transverse field} 
\label{sec:TransverseFieldExtra}

We derive the free energy and the effective Hamiltonian in the presence of a penalty transverse field. The total Hamiltonian we start from is Eq.~\eqref{full trans penalty hamil}.

We first succinctly reproduce (with some notational changes) the calculations in Section I of the Supplemental Material of Ref.~\cite{MNAL:15}, leading up to Eq.~(17) there. 
The partition function is computed by using the Suzuki-Trotter decomposition to separate the $\sigma^{z}$-only dependent part and the $\sigma^{x}$-only dependent part of the Hamiltonian $H$, i.e.,
$e^{-\beta(H^{x}+H^{z}) }=\lim_{M\to\infty}\left(e^{-{\beta\over M} H^{x}}e^{-{\beta\over M}  H^{z} }\right)^M$\,,
where
\bes
\begin{align}
H^z &= -{N}  \sum_{c=1}^{C}\left({1\over N}  \sum_{i=1}^{N}\sigma^{z}_{ic}   \right)^{p} -{\gamma}\sum_{c=1}^{C}\sum_{i=1}^{N} \sigma^{z}_{ic}\sigma^{z}_{i0}\\
H^x &= - \Gamma \sum_{c=1}^{C}\sum_{i=1}^{N} \sigma^{x}_{ic} -\epsilon C \Gamma \sum_{i=1}^{N} \sigma^{x}_{i0}\ .
\label{eq:H^x}
\end{align}
\ees

\onecolumngrid

In detail:
\begin{align}
Z_{M}&\equiv
\Tr \left(e^{-{\beta\over M} H^{x}}e^{-{\beta\over M}  H^{z} }\right)^M =\sum_{\{\sigma^{z}_{ic}, \sigma^{z}_{i0}   \}}
\langle \{ \sigma^{z}_{ic} \} |
 \left[
   \exp\left( -{\beta \over M}  H^z\right)
   \exp\left( -{\beta\over M} H^x\right)
 \right]^{M}
| \{ \sigma^{z}_{ic} \} \rangle \cr
&=\sum_{\{ \sigma^{z}_{ic},\sigma^{x}_{ic},\sigma^{z}_{i0}(\al)  \}}
\prod_{\al=1}^{M} 
\left\{ 
\exp\left( {\beta N\over M}  \sum_{c=1}^{C}\left({1\over N}  \sum_{i=1}^{N}\sigma^{z}_{ic}(\al)   \right)^{p} +{\beta\gamma\over M}\sum_{c=1}^{C}\sum_{i=1}^{N} \sigma^{z}_{ic}(\al)\sigma^{z}_{ic}(\al) + {\beta \Gamma\over M} \sum_{c=1}^{C}\sum_{i=1}^{N} \sigma^{x}_{ic}(\al) 
\right.\right.\cr
&
\left.\left. +{\beta\epsilon C \Gamma\over M}  \sum_{i=1}^{N} \sigma^{x}_{i0}(\al)
\right)
\prod_{i=1}^{N}\left(
\langle \sigma^{z}_{i0}(\al)|\sigma^{x}_{i0}(\al)\rangle \langle \sigma^{x}_{i0}(\al) | \sigma^{z}_{i0}(\al+1) \rangle
\prod_{c=1}^{C}
\langle \sigma^{z}_{ic}(\al)|\sigma^{x}_{ic}(\al)\rangle \langle \sigma^{x}_{ic}(\al) | \sigma^{z}_{ic}(\al+1) \rangle
\right)
\right\}\cr
&=\sum_{\{ \sigma^{z}_{ic},\sigma^{x}_{ic},\sigma^{z}_{i0}(\al)  \}} \prod_{\al=1}^{M} 
\left\{ \prod_{c=1}^{C}
 \int dm_{c\al} \int {d\tilde{m}_{c\al} \over 2\pi}
\exp\left({i\tilde{m}_{c\al}\left(m_{c\al}-{1\over N}  \sum_{i=1}^{N}\sigma^{z}_{ic} (\al)\right)}+{\beta N\over M} m_{c\al}^{p}  + \right.\right.\cr
&\left.+{\beta\gamma\over M}\sum_{i=1}^{N} \sigma^{z}_{ic}(\al)\sigma^{z}_{i0}(\al) + {\beta \Gamma\over M} \sum_{i=1}^{N} \sigma^{x}_{ic}(\al)
 +{\beta\epsilon \over M}  \sum_{i=1}^{N} \sigma^{x}_{i0}(\al)
\right)\times\cr
&
\left.
\prod_{i=1}^{N}\left(
\langle \sigma^{z}_{i0}(\al)|\sigma^{x}_{i0}(\al)\rangle \langle \sigma^{x}_{i0}(\al) | \sigma^{z}_{i0}(\al+1) \rangle
\prod_{c=1}^{C}
\langle \sigma^{z}_{ic}(\al)|\sigma^{x}_{ic}(\al)\rangle \langle \sigma^{x}_{ic}(\al) | \sigma^{z}_{ic}(\al+1) \rangle
\right)
\right\}\ .
\label{before static approximation}
\end{align}

By applying the static approximation $m_{c\alpha}=m_c,~\tilde{m}_{c\al}=\tilde{m}_{c}$, and changing variables $\tilde{m}\to { N\over M}\tilde{m}$, we obtain
\begin{align}
Z_{M}&=
\sum_{\{ \sigma^{z}_{ic},\sigma^{x}_{ic},\sigma^{z}_{i0}(\al)  \}}  \prod_{c=1}^{C}
\int dm_{c}  \int { d\tilde{m}_{c}\over 2\pi}
\exp\left(iM m_{c}\tilde{m}_{c}+{\beta N} m_{c}^{p} \right) \cr
&
\prod_{\al=1}^{M} \prod_{i=1}^{N} 
\left\{
\exp\left({-i{\tilde{m}_{c\al}\over N}  \sigma^{z}_{ic}(\al)}  
+{\beta\gamma\over M} \sigma^{z}_{ic}(\al)\sigma^{z}_{i0}(\al) + {\beta \Gamma\over M}  \sigma^{x}_{ic}(\al) 
+{\beta\epsilon \Gamma\over M} \sigma^{x}_{i0}(\al)
\right)  \right. \cr
&
\left.
\left(
\langle \sigma^{z}_{i0}(\al)|\sigma^{x}_{i0}(\al)\rangle \langle \sigma^{x}_{i0}(\al) | \sigma^{z}_{i0}(\al+1) \rangle
\prod_{c=1}^{C}
\langle \sigma^{z}_{ic}(\al)|\sigma^{x}_{ic}(\al)\rangle \langle \sigma^{x}_{ic}(\al) | \sigma^{z}_{ic}(\al+1) \rangle
\right)
\right\}\cr
&=
\sum_{\{ \sigma^{z}_{ic},\sigma^{x}_{ic},\sigma^{z}_{i0}(\al)  \}} 
\prod_{c=1}^{C}
\int dm_{c} \int {N d\tilde{m}_{c} \over 2\pi M}
\exp\left(iN m_{c}\tilde{m}_{c}+{\beta N} m_{c}^{p}\right) \cr
&
\prod_{\al=1}^{M} \prod_{i=1}^{N} 
\left\{ 
\exp\left({-i{\tilde{m}_{c}\over M}  \sigma^{z}_{ic}(\al)}  
+{\beta\gamma\over M} \sigma^{z}_{ic}(\al)\sigma^{z}_{i0}(\al) + {\beta \Gamma\over M}  \sigma^{x}_{ic}(\al) 
+{\beta\epsilon \Gamma\over M} \sigma^{x}_{i0}(\al)
\right)  \right. \cr
&
\left.
\left(
\langle \sigma^{z}_{i0}(\al)|\sigma^{x}_{i0}(\al)\rangle \langle \sigma^{x}_{i0}(\al) | \sigma^{z}_{i0}(\al+1) \rangle
\prod_{c=1}^{C}
\langle \sigma^{z}_{ic}(\al)|\sigma^{x}_{ic}(\al)\rangle \langle \sigma^{x}_{ic}(\al) | \sigma^{z}_{ic}(\al+1) \rangle
\right)
\right\} \ .
\label{eq:A3}
\end{align}
We take $M\to\infty$ to go from the classical system back to the quantum system:
\bea
Z
&=& \prod_{c=1}^{C}
\int dm_{c} \int d\tilde{m}_{c}
e^{(iN m_{c}\tilde{m}_{c}+{\beta N} m_{c}^{p}) }
\left(\Tr \prod_{c=1}^{C}e^{-i\tilde{m}_{c}\sigma^z_c+\beta\gamma
\sigma^{z}_{c}\sigma^{z}_{0}
+ {\beta \Gamma}  \sigma^{x}_{c}+{\beta\epsilon \Gamma} \sigma^{x}_{0}
}
\right)^{N}\ .
\label{partition with static}
\eea

This reproduces Eq.~(17) of the Supplemental Material of Ref.~\cite{MNAL:15}. Note that here we did not divide the penalty transverse field term by $C$ (denoted $K$ there). 

The saddle point equation for $m_{c}$ is:
\bea
i\tilde{m}_{c}+\beta p m^{p-1}_{c}=0\ ,
\label{mtilde and m}
\eea
which gives
\begin{align}
Z &= \prod_{c}^{C}
\int dm_{c} 
e^{(\beta N (1-p)m_{c}^{p}) }
\left(\Tr \prod_{c=1}^{C}e^{\beta (p m^{{p-1}}_c\sigma^z_c+\gamma
\sigma^{z}_{c}\sigma^{z}_{0}
+ { \Gamma}  \sigma^{x}_{c}+{\epsilon \Gamma} \sigma^{x}_{0})
}
\right)^{N}\ .
\label{static-partition}
\end{align}

This yields the free energy $F$, defined by
$Z=\exp(-\beta NC F)$, as in Eq.~\eqref{trans penalty partition func}, i.e., 
$F{/C}=(p-1)m^{p}-{1\over \beta C}\log \Tr\prod_{c=1}^{C}e^{-\beta H_{\mathrm{eff}}}$, where $H_{\mathrm{eff}}=-(p m^{{p-1}}\sigma^{z}_{c}+\gamma
\sigma^{z}_{c}\sigma^{z}_{0}
+ { \Gamma}  \sigma^{x}_{c}+{\epsilon \Gamma } \sigma^{x}_{0})$ is the effective Hamiltonian [Eq.~\eqref{eq:Heff}], 
which describes the fluctuation around the mean field value.

Note that Ref.~\cite{MNAL:15} did not pursue the effect of the penalty transverse field [$\epsilon$ was set to zero after Eq.~(17) there] and was concerned primarily with the large $\beta$ limit, which we do not assume here.

When $\epsilon=0$, copies with different values of $c$ are decoupled.
Therefore, one can obtain the free energy analytically.
Consider the trace: 
\begin{align}
&\Tr \prod_{c=1}^{C}e^{\beta (p m^{p-1}_c\sigma^z_c+\gamma
\sigma^{z}_{c}\sigma^{z}_{0}
+ { \Gamma}  \sigma^{x}_{c})
}
=\Tr \prod_{c=1}^{C}e^{\beta (p m^{p-1}_c\sigma^z_c+\gamma
\sigma^{z}_{c}
+ { \Gamma}  \sigma^{x}_{c})}
+
\Tr \prod_{c=1}^{C}e^{\beta (p m^{p-1}_c\sigma^z_c-\gamma
\sigma^{z}_{c}
+ { \Gamma}  \sigma^{x}_{c})}
\cr
&\qquad =
\prod_{c=1}^{C}\left(e^{\beta\sqrt{(p m^{p-1}_c+\gamma)^2+\Gamma^2}}+e^{-\beta\sqrt{(p m^{p-1}_c+\gamma)^2+\Gamma^2}}
\right) 
+\prod_{c=1}^{C}\left(e^{\beta\sqrt{(p m^{p-1}_c-\gamma)^2+\Gamma^2}}+e^{-\beta\sqrt{(p m^{p-1}_c-\gamma)^2+\Gamma^2}}
\right)\cr
&\qquad =\sum_{s\in{\{-,+\}}}\prod_{c=1}^{C}[2\cosh(\beta Q_{s,c})] \ .
\end{align}
Then the free energy is
\bea
&&F/C={1\over C}\sum_{c=1}^{C}(p-1)m_c^{p}-{1\over \beta C}\log \sum_{s\in{\{-,+\}}}\prod_{c=1}^{C}[2\cosh(\beta Q_{s,c})].\cr
&&
\label{free energy appendix}
\eea
\twocolumngrid

If we set $m_c\equiv m$ and $Q_{s,c}\equiv Q_{s}$, this recovers Eq.~\eqref{free energy mean field}.
The saddle point equation~\eqref{saddle point eq0} is obtained by evaluating
$\pd F/\pd m_c=0$.


\section{Instantons, tunneling, and a relation between free energy and gaps} 
\label{app:Instanton}

\begin{figure*}[ht]
\begin{tabular}{ccc}
\subfigure[]{   \includegraphics[scale=0.45]{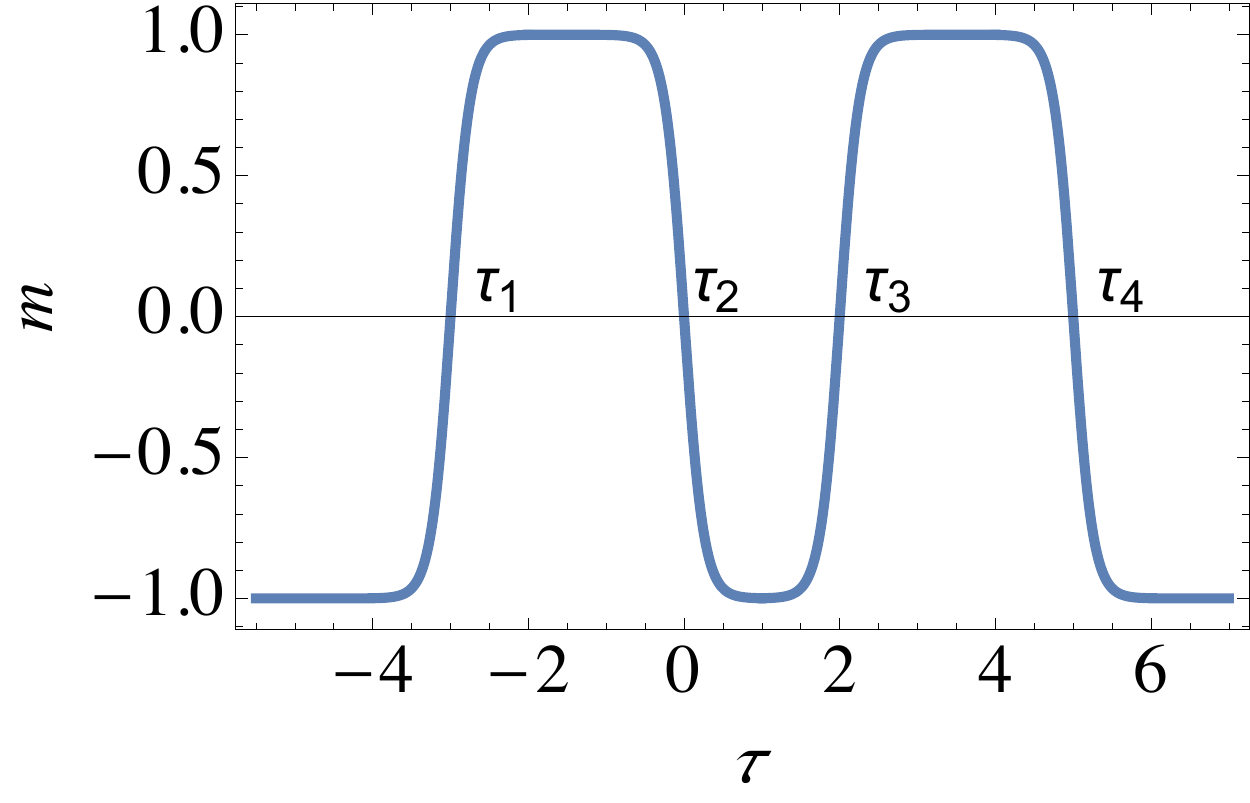} \label{instansol} }  &  
\subfigure[]{ \includegraphics[scale=0.45]{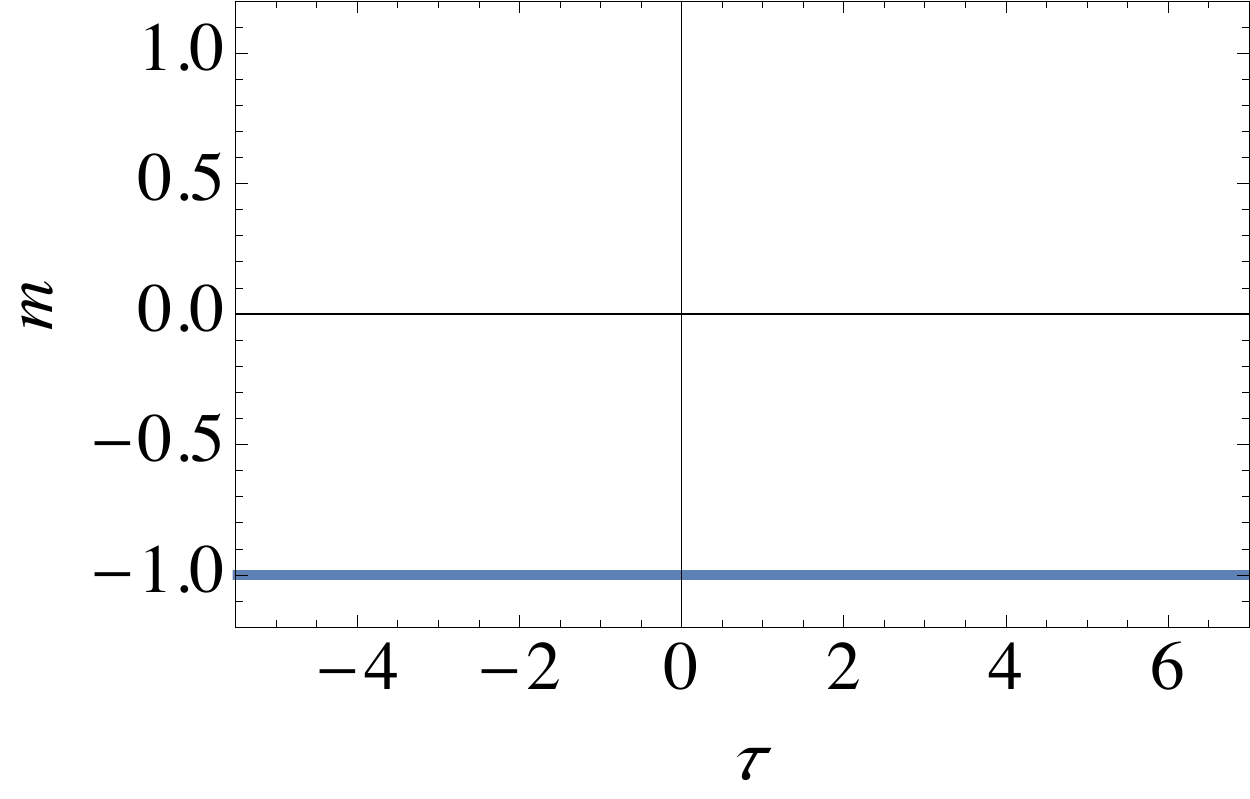}   \label{zeroinst}} & 
\subfigure[]{     \includegraphics[scale=0.45]{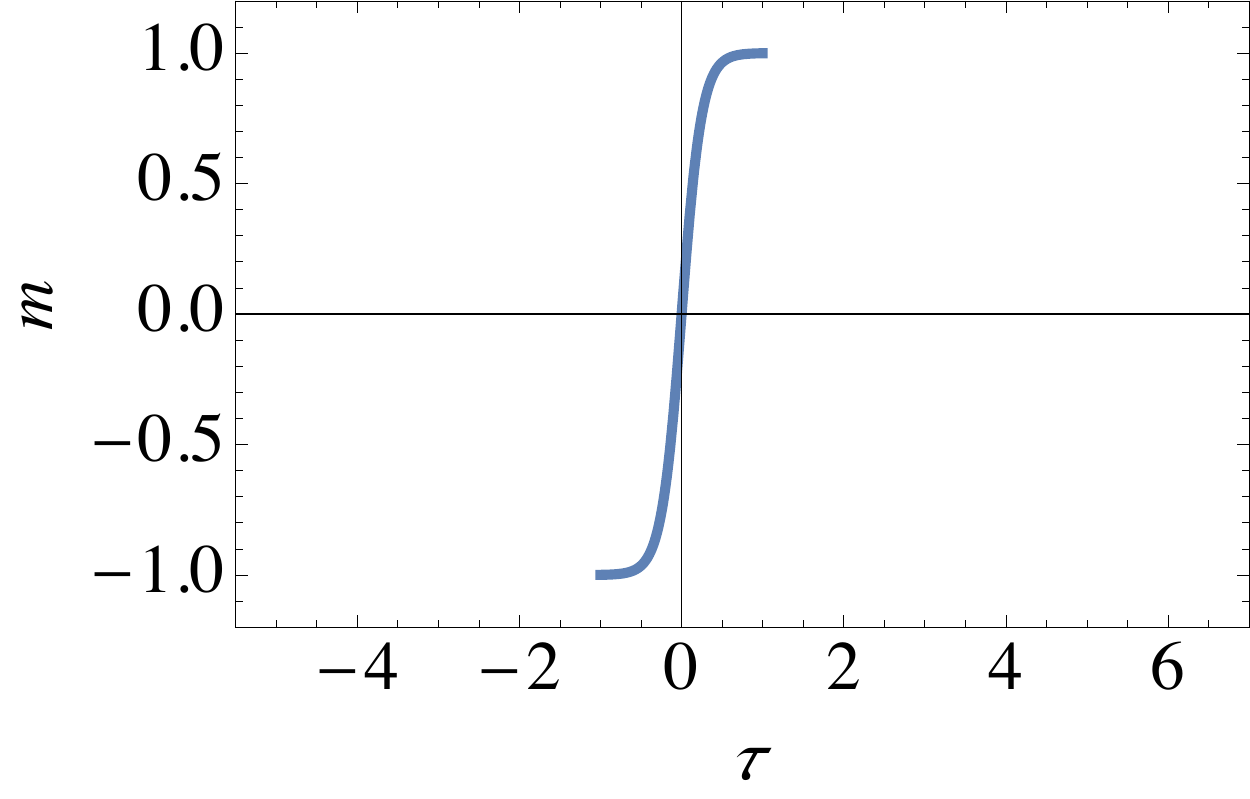}   \label{oneinst}} 
\end{tabular}     
\caption{Illustration of instanton trajectories. (a) the four-instanton solution. If the transition between two minima happens very quickly, the Euclidean action can be approximated by the sum of the zero instanton action shown in (b)
and four times the transition action shown in (c). The partition function is multiplied by $\beta^4/4!$ due to the possible locations of $\tau_{i}$, $i\in\{1,2,3,4\}$.}
\label{fig:instantons}
\end{figure*}

In the main text, we argued that the width and height of the free energy barrier at the quantum phase transition point can be used to estimate the ground state energy gap. In this section, we provide  support for this argument. We closely follow (and expand upon) Ref.~\cite{Jorg:2010qa}, which showed that the energy gap between the ground state and the first excited state comes from the tunneling between the two free energy minima.  The tunneling rate is given by an instanton solution, which is a solution of the Euclidean action connecting the two free energy minima.  In our derivation of the free energy in Appendix ~\ref{sec:TransverseFieldExtra}, we neglected such solutions by using the static approximation, which assumes that the order parameter $m$ is constant along the Euclidean time direction $\tau$.  Therefore, to get an estimate of the energy gap, we must relax the static approximation.

Let us assume that we have fixed the value $\Gamma$ such that the saddle-point free energy has two degenerate minima at $m_1$ and $m_2$ with value $F_0$.  These two minima contribute $2 e^{-\beta N F_0}$ to the partition function $Z$.  We wish to consider non-perturbative corrections to $Z$ by relaxing the static approximation and integrating $m$ over all possible configurations along  $\tau$ satisfying the periodic boundary condition $m(\tau =0) = m(\tau = \beta)$.

The partition function [Eq.~\eqref{static-partition}] and the free energy [Eq.~\eqref{free energy appendix}] are obtained by assuming that $m(\tau)$ is a constant.
We therefore return to Eq.~\eqref{before static approximation} and proceed without the static approximation.  For simplicity, we set the penalty transverse field to zero, i.e., we take $\epsilon=0$.  We first perform the sum over the penalty qubit values, which yields: 
\begin{widetext}
\begin{align}
Z_M &= \sum_{\{ \sigma^{z}_{ic},\sigma^{x}_{ic}(\al)  \}} \prod_{\al=1}^{M} 
\left\{ \prod_{c=1}^{C}
 \int dm_{c\al} \int {d\tilde{m}_{c\al} \over 2\pi}
\sum_{s_0 = \left\{+1,-1 \right\}} \exp\left({i\tilde{m}_{c\al}\left(m_{c\al}-{1\over N}  \sum_{i=1}^{N}\sigma^{z}_{ic} (\al)\right)}+{\beta N\over M} m_{c\al}^{p}  + \right.\right.\cr
&\left.+{s_0 \beta\gamma\over M}\sum_{i=1}^{N} \sigma^{z}_{ic}(\al) + {\beta \Gamma\over M} \sum_{i=1}^{N} \sigma^{x}_{ic}(\al)
\right)\times
\prod_{i=1}^{N}\left(
\prod_{c=1}^{C}
\langle \sigma^{z}_{ic}(\al)|\sigma^{x}_{ic}(\al)\rangle \langle \sigma^{x}_{ic}(\al) | \sigma^{z}_{ic}(\al+1) \rangle
\right) \ .
\end{align}
\end{widetext}
After changing variables $\tilde{m}_{c\alpha}\to { N\over M}\tilde{m}_{c\alpha}$, a variation with respect to $m_{c \alpha}$ gives the on-shell value of  $\tilde{m}_{c\alpha}$
\beq
i \tilde{m}_{c \alpha} = - \beta p m_{c \alpha}^{p-1} \ ,
\eeq
which we can then insert into our expression for the partition function to give in the $M \to \infty$ limit:
\begin{align}
Z&=\prod_{c=1}^{C}\int_{m(0) = m(\beta)} \hspace{-1.25cm} \mathcal{D}m_c(\tau)
\exp\left(
N\int_{0}^{\beta}d\tau (1-p)m^{p}_{c}(\tau)
\right)\times \cr
&
\left(\Tr \prod_{c=1}^{C}e^{\int_{0}^{\beta} d\tau(p m^{{p-1}}_c(\tau)\sigma^z_c(\tau)+\gamma
\sigma^{z}_{c}(\tau)
+ { \Gamma}  \sigma^{x}_{c} (\tau))
}\right. \cr
&+\left.
\Tr \prod_{c=1}^{C}e^{\int_{0}^{\beta} d\tau(p m^{{p-1}}_c(\tau)\sigma^z_c(\tau)-\gamma
\sigma^{z}_{c}(\tau)
+ { \Gamma}  \sigma^{x}_{c}(\tau))
}
\right)^{N} \ ,
\label{exact partition}
\end{align}
where we have expressed:
\begin{align}
\lim_{M \to \infty} \prod_{\alpha = 1}^M e^{\frac{\beta}{M} \left( \Gamma \sigma^x + p m^{p-1} \sigma^z \pm \gamma \sigma^z \right)} \nonumber \\
=  e^{\int_0^\beta d \tau \left( \Gamma \sigma^x(\tau) + p m^{p-1} \sigma^z(\tau) \pm \gamma \sigma^z(\tau) \right) }
\end{align} 
This is a natural generalization of the result in Ref.~\cite{Jorg:2010qa} for the ferromagnetic $p$-model:
\bea
Z=\int_{m(0)=m(\beta)} \hspace{-1.25cm} \mathcal{D}m(\tau)e^{-S[m(\tau)]} \ ,
\label{def of action}
\eea
with the action
\begin{align} 
\label{eqt:JorgAction}
S[m(\tau)]= 
-&N \int_0^{\beta} d \tau \left( (p-1) m(\tau)^p \right)  
\nonumber \\
+&  N \log \Tr e^{\int_0^\beta d\tau \left( \Gamma \sigma^x(\tau) + p m^{p-1} \sigma^z(\tau) \right) }
\ .
\end{align}
%
%
There are many classical solutions (solutions of the equation of motion derived from the Euclidean action $S[m(\tau)]$) that satisfy the boundary condition $m(0)=m(\beta)$.
The simplest solutions are $m(\tau)=m_1$ and $m(\tau)=m_2$ for all $0\le\tau\le\beta$.
These are zero instanton solutions, and the Euclidean action takes the value $S_0=\beta  N F_0$ for each solution, resulting in the contribution $2 e^{-S_0}$ to the partition function. More generally, we can consider instanton solutions that perform an even number (due to the periodic boundary condition) of discrete jumps between $m_1$ and $m_2$ and their contribution to the partition function:
\beq
Z = \sum_{k=0}^{\infty} Z_{2k}\ .
\eeq

The next simplest solutions after the zero-instanton solutions are the two-instanton solutions. These start with $m(0)=m_{1(2)}$,
stay in the same minimum, then switch to $m(\tau_1)=m_{2(1)}$ at $\tau=\tau_1$, then stay in the same minimum until $\tau=\tau_2$, and then switch to the initial minimum $m(\tau_2)=m_{1(2)}$.
If we assume that these transitions happen almost instantly, 
the two-instanton Euclidean action can be approximated by the sum of the zero-instanton action $S_0$ and the sum of the two transition actions $S_{\text{trans}}$, i.e., $S_2 \approx S_0 + 2S_{\text{trans}}$.
The partition function (due to the path integral) must account for the possible locations of $\tau_{1}$ and $\tau_2$, namely 
\beq
\int_{0}^{\beta}d\tau_2\int_{0}^{\tau_2}d\tau_1 = \beta^2/2! \ .
\eeq
Thus, we have for the two instanton partition function:
\beq 
\label{eqt:Z2}
Z_2 = \frac{2 \beta^2}{2!} e^{-S_2} = \varepsilon^2 \beta^2 e^{-S_0} \ ,
\eeq
where 
\beq \label{eqt:epsilon}
\varepsilon \equiv e^{-S_{\text{trans}}}\ .
\eeq

If we consider the case of $2 k$ transitions occurring at $\tau_1 < \tau_2 <\dots <\tau_{2k}$, illustrated in Fig.~\ref{fig:instantons} for $k=2$, then we have 
\begin{align}
Z_{2k} &= 2 e^{-S_0} \varepsilon^{2k}  \int_0^{\beta}  d\tau_{2k}  \int_0^{\tau_{2k}} d \tau_{2k-1} \dots \int_0^{\tau_{2}} d \tau_1  \notag \\
&=  2 e^{-\beta N F_0} \frac{(\beta \varepsilon)^{2k}}{(2k)!}\ ,
\end{align}
where for simplicity we assumed that all transitions are equal. 
This then gives:
\bes
\begin{align}
Z &= 2 e^{-\beta N F_0} \cosh( \beta  \varepsilon) = e^{- \beta(N F_0 + \varepsilon)} + e^{- \beta( NF_0 - \varepsilon)} \\
&= \Tr[\exp(-\beta \tilde{H})]\ , 
\end{align}
\ees
where
\beq
\qquad \tilde{H} = 
\left(
\begin{array}{cc}
 N F_0   & \varepsilon   \\
   \varepsilon  & N F_0  
\end{array}
\right) \ .
\eeq
We see that this effectively describes a two-level system with an energy gap $\tilde{\Delta} = 2 \varepsilon$, i.e., the energy gap is related to the transition from $m_1$ to $m_2$, which is the one-instanton contribution:
\beq
{Z(\text{one-instanton})\over Z(\text{zero-instanton})} = \frac{2 e^{-\beta  N F_0} \beta \varepsilon}{2 e^{-\beta N F_0}} = \beta \varepsilon\ .
\eeq
Moreover, the gap of the effective Hamiltonian $\tilde{H}$ approaches the gap $\Delta$ of the QA (closed system) Hamiltonian in the $\beta\to \infty$ limit. This is the sense in which the instantonic approach and tunneling between valleys of the free energy allows us to estimate the gap of the QA Hamiltonian. 

It remains to calculate $S_{\text{trans}}$. We can do this 
with a two-instanton ans{\"a}tz for the ferromagnetic $p$-model where sharp transitions occur at  $\tau = s$ and $\tau = \beta$:
\beq
m(\tau) = \left\{ \begin{array} {lr}
 m_1 \ , & 0 \leq\tau < s \\
 m_2 \ , &   s \leq \tau < \beta
\end{array} \right. \ ,
\eeq
where $m(0) = m(\beta)$.  While this is not technically a two-instanton solution, i.e., it is not a solution to the equations of motion derived from the action in Eq.~\eqref{eqt:JorgAction}, it gives a (crude) upper-bound on the two-instanton 
action since a true two-instanton solution would minimize the action.  For this ans{\"a}tz we have:
\begin{figure*}[ht]
\subfigure[]{\includegraphics[width=0.6\columnwidth]{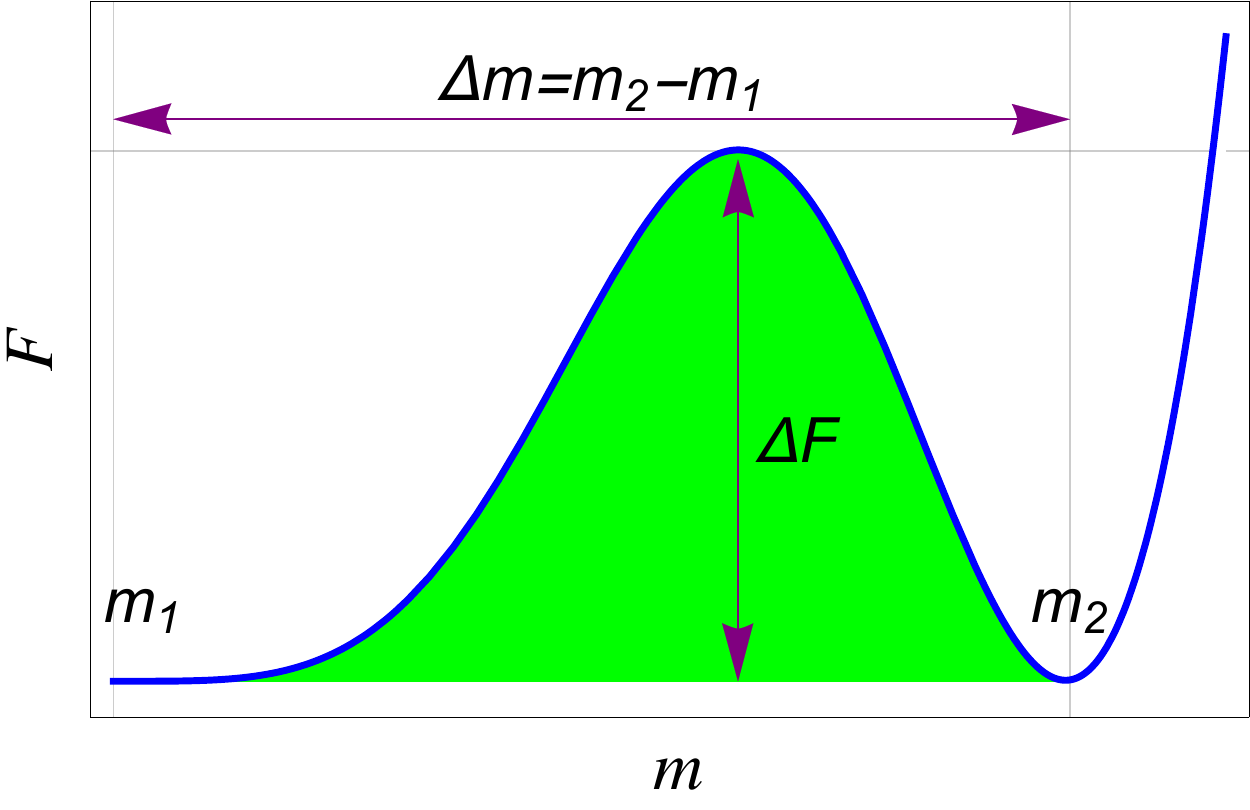} \label{barrierarea}}
   \subfigure[]{\includegraphics[width=0.6\columnwidth]{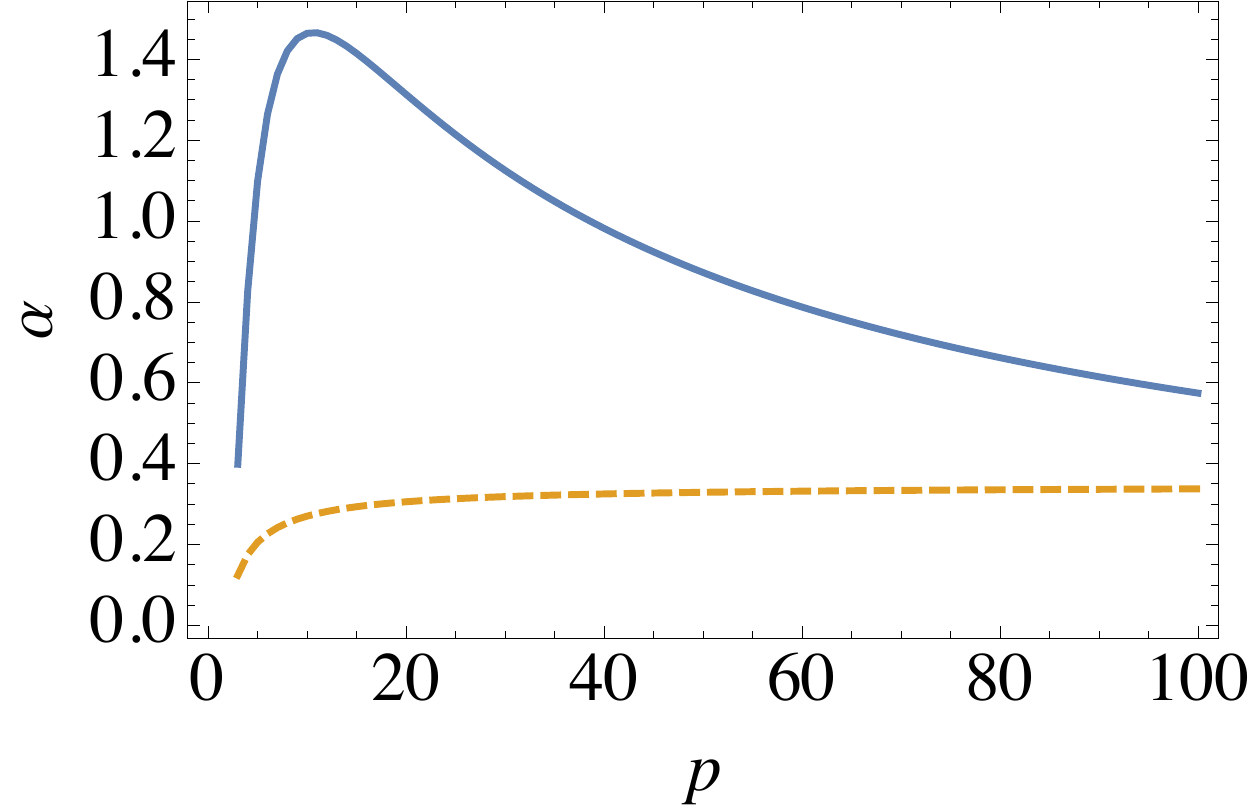} \label{fig:ActionComparison1}}
   \subfigure[]{\includegraphics[width=0.6\columnwidth]{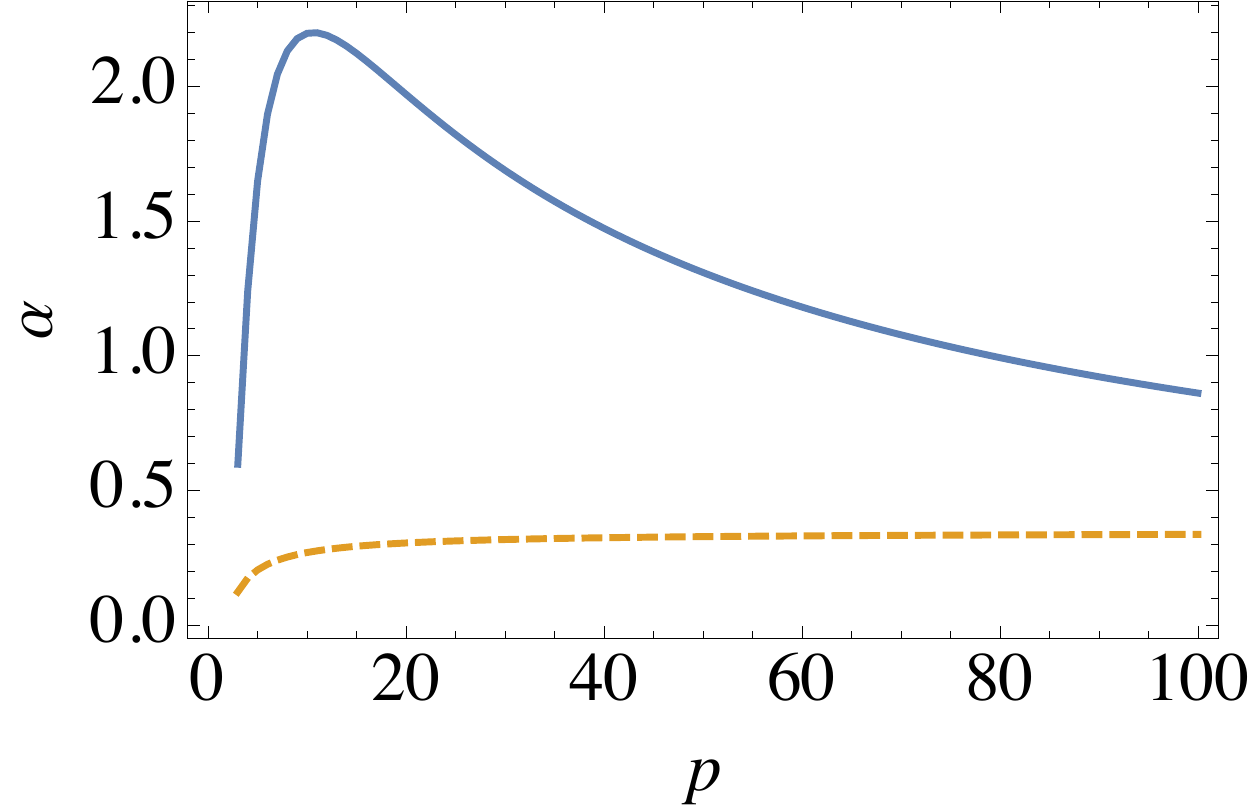} \label{fig:ActionComparison2} }
\caption{ A comparison of the gap scaling coefficient $\Delta = e^{-\alpha N}$ estimated using the sharp instanton and using the area under the free energy potential barrier for the ferromagnetic $p$-model.  (a) The area of the potential barrier (shaded region). (b) The comparison for $\beta = 20$.  (c) The comparison for $\beta = 30$.  The solid blue curve corresponds to the area under the free energy barrier [Eq.~\eqref{eqt:area}], while the dashed orange curve is the sharp instanton coefficient $-  \log   \left| \braket{\lambda_+(m_1)}{\lambda_+(m_2)} \right|)$.}
\end{figure*}
\begin{align}
 \prod_{\alpha = 1}^M e^{\frac{\beta}{M} \left( \Gamma \sigma^x + p m^{p-1} \sigma^z \right)} &= \prod_{\alpha=1}^{M_s} e^{\frac{\beta}{M} \left( \Gamma \sigma^x + p m_1^{p-1} \sigma^z \right)}  \nonumber \\
 & \times \prod_{\alpha=M_s + 1}^{M} e^{\frac{\beta}{M} \left( \Gamma \sigma^x + p m_2^{p-1} \sigma^z \right)} \ ,
\end{align}
where we have defined $M_s$ such that $\lim_{M \to \infty} \beta M_s / M = s$.   We can calculate the trace of this operator by diagonalizing each term:
\begin{widetext}
\label{eqt:traceTerms}
\begin{align} 
&\Tr \lim_{M \to \infty} \prod_{\alpha = 1}^M e^{\frac{\beta}{M} \left( \Gamma \sigma^x + p m^{p-1} \sigma^z \right)} \notag  \\
&\quad 
= \sum_{r\in\{-,+\}}\bra{\lambda_r(m_1)} \lim_{M \to \infty} 
\left[ \prod_{\alpha=1}^{M_s} \left(e^{\lambda_+(m_1)} \ketbra{\lambda_+(m_1)}{\lambda_+(m_1)} 
 \right. 
+ e^{\lambda_-(m_1)} \ketbra{\lambda_-(m_1)}{\lambda_-(m_1)} \right)
\times \nonumber \\ 
&\qquad  
\prod_{\alpha=M_s + 1}^{M} \left(e^{\lambda(m_2)} \ketbra{\lambda_+(m_2)}{\lambda_+(m_2)} 
\left. 
+ e^{-\lambda(m_2)} \ketbra{\lambda_-(m_2)}{\lambda_-(m_2)} \right) \right] \ket{\lambda_r(m_1)}\ ,
\end{align}
\end{widetext}
where we have denoted the two eigenvalues by $\lambda_{\pm}(m) = \pm \lambda(m)$, where $ \lambda(m)= \sqrt{ \Gamma^2 + (p m^{p-1})^2 }$, with the respective orthonormal eigenvectors:
\begin{align}
\ket{\lambda_\pm(m)} & = \frac{1}{\sqrt{ \Gamma^2 + (p m^{p-1} \pm \lambda(m))^2}} \nonumber \\
& \times \left[ \left(p m^{p-1} \pm \lambda(m) \right) \ket{0} + \Gamma \ket{1} \right]  \ ,
\end{align}
and used that fact that we are free to choose any fixed orthonormal basis to compute the trace (we selected $\ket{\lambda_\pm(m_1)}$). 
Interchanging the order of the sum and product we thus obtain:
\begin{align}
&\Tr \lim_{M \to \infty} \prod_{\alpha = 1}^M e^{\frac{\beta}{M} \left( \Gamma \sigma^x + p m^{p-1} \sigma^z \right)}\\
&\quad =e^{ s \lambda(m_1)} e^{(\beta-s) \lambda(m_2)} \left| \braket{\lambda_+(m_1)}{\lambda_+(m_2)} \right|^2 \nonumber  \\
&\qquad+ e^{s \lambda(m_1)} e^{-(\beta-s) \lambda(m_2)} \left| \braket{\lambda_+(m_1)}{\lambda_-(m_2)} \right|^2  \nonumber  \\
&\qquad+ e^{- s \lambda(m_1)} e^{(\beta - s) \lambda(m_2)} \left| \braket{\lambda_-(m_1)}{\lambda_+(m_2)} \right|^2  \nonumber  \\
&\qquad+ e^{- s \lambda(m_1)} e^{-(\beta - s) \lambda(m_2)} \left| \braket{\lambda_-(m_1)}{\lambda_-(m_2)} \right|^2  \nonumber \ .
\end{align}

Let us now consider the case where $\beta \gg 1$ such that the first term in Eq.~\eqref{eqt:traceTerms} dominates.  We then have for this particular sharp two-instanton action, by combining Eqs.~\eqref{eqt:JorgAction}, \eqref{eqt:Z2}, and \eqref{eqt:traceTerms}:
\begin{align}
S_{2}/N &= s (p-1) m_1^p + (\beta - s) (p-1) m_2^p - s \lambda(m_1) \nonumber \\
& -(\beta - s) \lambda(m_2) - 2 \log   \left| \braket{\lambda_+(m_1)}{\lambda_+(m_2)} \right|  .
\end{align}
Because we have assumed that we are calculating the instanton at the point where the free energy $F(m)$ is degenerate, i.e.,
\bes
\begin{align}
&F(m_1) = (p-1)m_1^p - \lambda(m_1) = \\
&F(m_2) = (p-1)m_2^p - \lambda(m_2) \ ,
\end{align}
\ees
we can write our sharp two-instanton action in an $s$-independent way as:
\beq
S_2  =   N \beta F(m_1) -  2 N \log   \left| \braket{\lambda_+(m_1)}{\lambda_+(m_2)} \right| \ .
\eeq
Therefore, we can estimate the two-instanton partition function, using Eq.~\eqref{eqt:Z2}, as:
\beq \label{eqt:Z2b}
Z_2 = \beta^2 e^{-S_2} = \beta^2 e^{- S_0} e^{-2 N \log   \left| \braket{\lambda_+(m_1)}{\lambda_+(m_2)} \right|} \ ,
\eeq
where we have used the zero-instanton action $S_0 = N \beta F(m_1)$.  Comparing with the expression for $Z_2$ in Eq.~\eqref{eqt:Z2}, we can now readily identify a lower bound on the quantity $\varepsilon$:
\beq
\varepsilon \geq e^{-N(- \log   \left| \braket{\lambda_+(m_1)}{\lambda_+(m_2)} \right|)} \ .
\eeq
Recall that the reason that the sharp instanton only gives a lower bound on the gap is that it is not a true solution to the equations of motion. Therefore the coefficient $-2 N  \log   \left| \braket{\lambda_+(m_1)}{\lambda_+(m_2)} \right|$ provides an upper bound on the coefficient of the scaling of the gap.  This was confirmed in Ref.~\cite{Jorg:2010qa}.
Therefore, the gap is lower bounded by
\beq
\Delta = 2 \varepsilon \geq   2 e^{ - N (-  \log   \left| \braket{\lambda_+(m_1)}{\lambda_+(m_2)} \right|) } \ .
\eeq
This agrees with the result in Ref.~\cite{Jorg:2010qa} for the sharp instanton.

We now wish to see whether we can further lower-bound this estimate by 
\beq \label{eqt:area}
e^{- \beta N \int_{m_1}^{m_2} \left[F(m) - F(m_1)\right] dm} \ ,
\eeq
an action term corresponding to the area under the free energy potential barrier, as depicted in Fig.~\ref{barrierarea}.  In the case of the ferromagnetic $p$-model, $m_1 = 0$, and we can numerically solve for $m_2$.  We show in Figs.~\ref{fig:ActionComparison1} and \ref{fig:ActionComparison2} that $\beta \int_{m_1}^{m_2} [F(m) - F(m_1)] >  - \log   \left| \braket{\lambda_+(m_1)}{\lambda_+(m_2)} \right| )$,  for sufficiently small $p$, and hence it provides an upper bound for the sharp instanton prediction (and hence a lower bound on the gap) in this regime.  As we increase $\beta$ the regime of $p$ where this is valid grows.  However, as we increase $\beta$, the bound is generally less tight. Therefore, we may conclude that $\Delta \gtrsim e^{-\Delta m \Delta F N}$, which is Eq.~\eqref{eq:gap-F}. The simple ferromagnetic $p$-spin model was used here to illustrate the idea behind the justification of Eq.~\eqref{eq:gap-F}, but qualitatively the situation would not be very different for the case with a penalty term.

Before closing this section, we mention that we can explicitly have a time-derivative term for $m(\tau)$ in the action [Eqs.~\eqref{exact partition} and \eqref{eqt:JorgAction}] by diagonalizing the instantaneous Hamiltonian. This method was used in Ref.~\cite{Knysh:2016iq} in a similar setup.
For the above sharp instanton ans{\"a}tz, this derivative term determines the value of $S_{\text{trans}}$. 
More generally one can solve the equation of motion and evaluate the Euclidean action with it.

%

\section{Free energy diagrams} \label{sec:extraF}
In order to better understand the phase diagram associated with $p>2$, we consider the following three cases: 
(i) a low temperature and $\gamma<\gamma_c$, $(T,\gamma)=(0.025,0.5)$;
(ii) a low temperature and  $\gamma\simeq\gamma_c$, $(T,\gamma)=(0.025,0.8)$;
(iii) a high temperature and $\gamma<\gamma_c$,  $(T,\gamma)=(0.1,0.7)$.

\subsubsection{Case (i): low temperature and $\gamma<\gamma_c$}
Let us first consider case (i). The free energy  when $\gamma=0.5$  and $T = 0.025$  is shown in Fig.~\ref{p4T0025g05all}  for various values of $\Gamma$. When quantum fluctuations are large ($\Gamma=1.95$), the unique minimum is at $m=0$.
As $\Gamma$ decreases, the free energy forms a new local minimum around $m=m_{\text{small}}\simeq0.3$ at $\Gamma=1.91$ [Fig.~\ref{p4T0025g05G191}].
At this point there is a first order phase transition from $m=0$ to $m_{\text{small}}$.
As $\Gamma$ decreases further, there is another first order phase transition at $\Gamma=1.846$ [Fig.~\ref{p4T0025g05G1846}].
The free energy minimum changes from $m=m_{\text{small}}\simeq0.328$ to $m=m_{\text{large}}\simeq 0.844$.
Below this value of $\Gamma$, the free energy minimum stays at $m_{\text{large}}$. Note that $m=0$ is always an unstable point at zero temperature and for any finite value of $\gamma$, and there is only a single first order phase transition at $\Gamma_c=1.847$ \cite{MNAL:15}  from $\ms$ to $m_{\text{large}}$. 

The phase transition from $m=0$ to $\ms$ at finite temperature is governed by a very small free energy barrier. The width and height of the potential barrier in the second transition, from $\ms$ to $m_{\text{large}}$, are much larger than those of the $m=0$ to $\ms$ transition. 
Since the tunneling rate depends on the height and width of the free energy barrier, this second transition dominates. Nevertheless, the potential barrier is smaller than in the case of a single phase transition from $m=0$ to $m_{\text{large}}$. The comparison of the free energies at the
phase transitions  for $\gamma=0$ (between $m=0$ and $m_{\text{large}}$)
 and for $\gamma=0.5$ (between $m=m_{\text{small}}$ and $m_{\text{large}}$)
  at $T=0.025$ is
shown in Fig.~\ref{p4T0025gcomp}.   
The introduction of the penalty terms thus results in the breaking up of a single, large free energy barrier into multiple smaller ones. In this sense, the penalty term weakens the phase transition. We expect that having multiple first order phase transitions from $m=0$ to $\ml$ results in more efficient QA than a single first order phase transition. 

This beneficial splitting of the phase transition, however, occurs only for sufficiently small values of the temperature.  As the temperature increases, $\ms$ approaches  $0$ and above a certain ``branching point" temperature there is only one phase transition from $m=0$ to $m_{\text{large}}$. Thus, above the branching point, the first order phase transition is not split.

\subsubsection{Case (ii): low temperature and $\gamma\simeq\gamma_c$}

Next, we consider case (ii), with $(T,\gamma)=(0.025,0.8)$. The chosen value of $\gamma$ corresponds to the point where  the first order phase transition disappears at zero temperature. The free energy for various values of $\Gamma$ is shown in Fig.~\ref{p4T0025g08all}.
When $\Gamma$ is very large, there is a unique ground state at $m=0$.
As $\Gamma$ becomes smaller, a new local minimum appears around $m=m_{\text{small}}\simeq0.3$ and there is a first order transition 
between $m=0$ and  $m_{\text{small}}$ [Fig.~\ref{p4T0025g08G295}].
As $\Gamma$ decreases further, the minimum of the free energy becomes almost flat between $\ms \simeq 0.5$ and $\ml \simeq 0.7$ [Fig.~\ref{p4T0025g08G222}].
Therefore the state can shift (almost) smoothly from the local minimum $m_{\text{small}}$ to the global minimum $m_{\text{large}}$.
For $\gamma\le\gamma_c$, the first order phase transition between $\ms$ and $\ml$ was a remnant of that at zero temperature.
As this first order phase transition disappears for $\gamma\ge \gamma_c$ at $T=0$, the corresponding phase transition also disappear
at finite but low temperature.

\subsubsection{Case (iii): high temperature and $\gamma<\gamma_c$}

\begin{figure*}[ht]
\begin{tabular}{ccc}
\subfigure[]{     \includegraphics[scale=0.6]{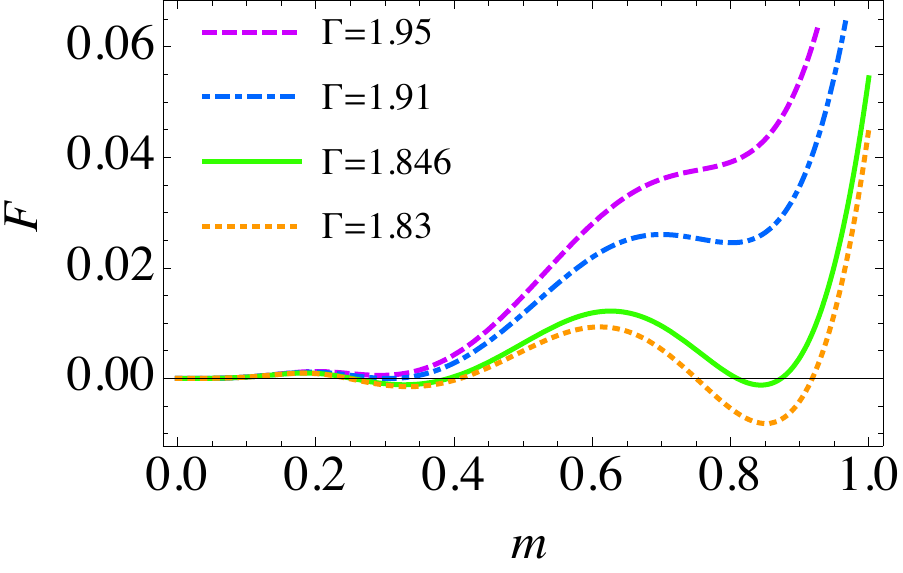} \label{p4T0025g05all}  }  &
\subfigure[]{     \includegraphics[scale=0.6]{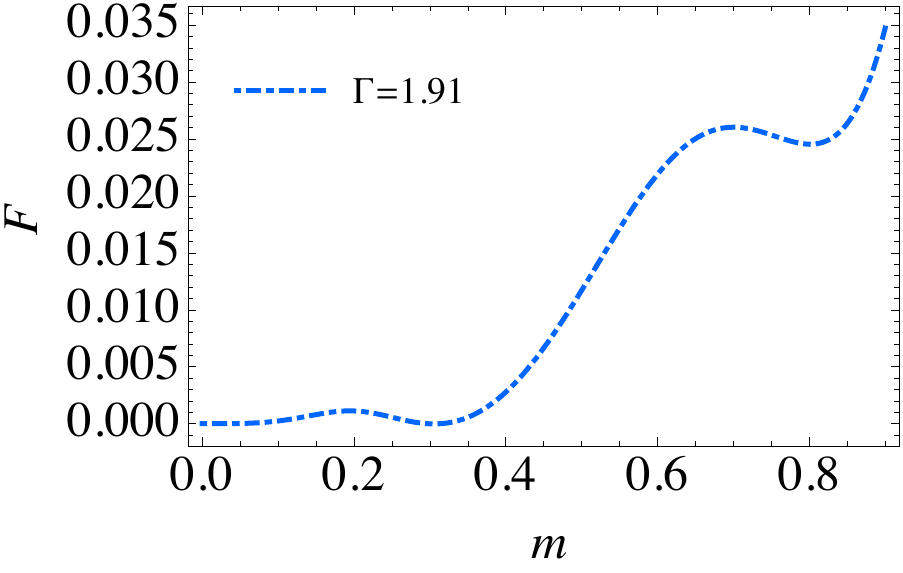}   \label{p4T0025g05G191} }  &
\subfigure[]{        \includegraphics[scale=0.6]{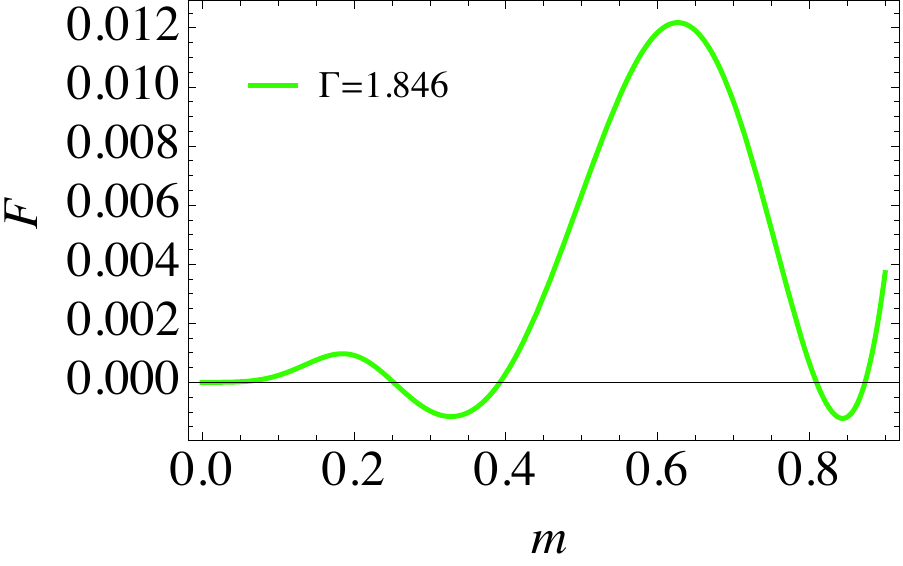}  \label{p4T0025g05G1846}  }   \\
\subfigure[]{     \includegraphics[scale=0.6]{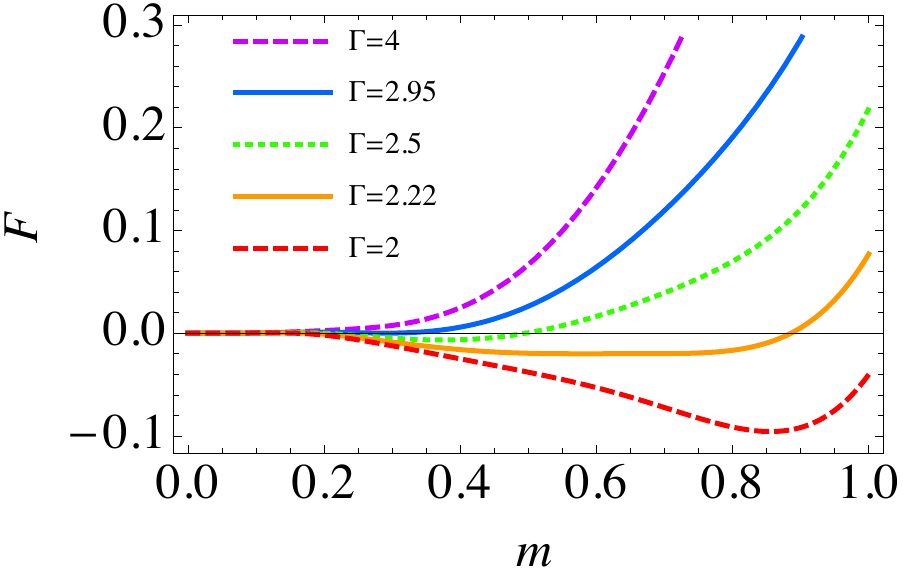}  \label{p4T0025g08all} }  &
\subfigure[]{     \includegraphics[scale=0.6]{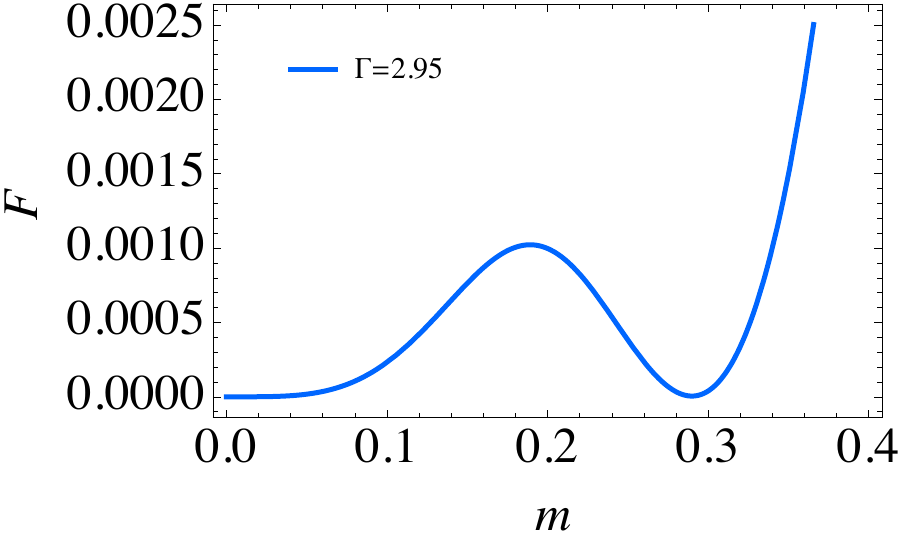} \label{p4T0025g08G295} } &
\subfigure[]{        \includegraphics[scale=0.6]{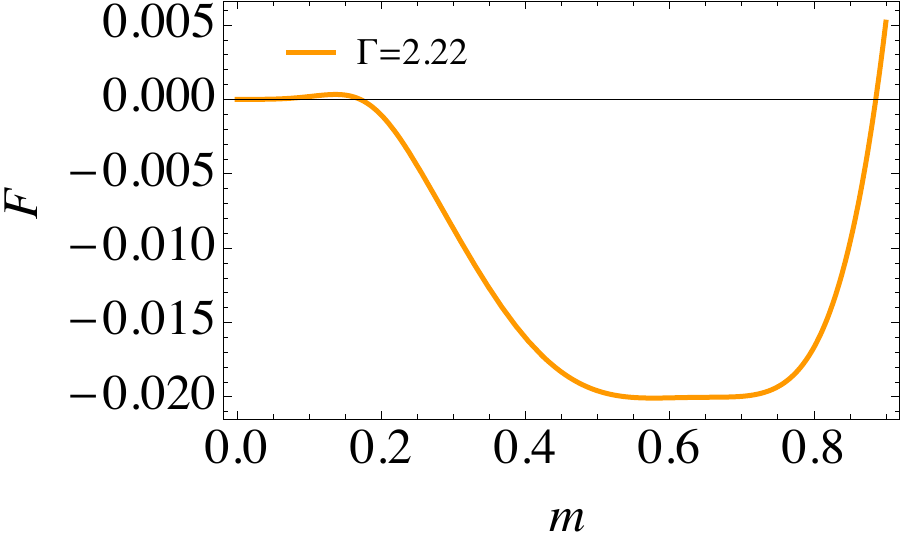}  \label{p4T0025g08G222} }   \\
 \subfigure[]{     \includegraphics[scale=0.6]{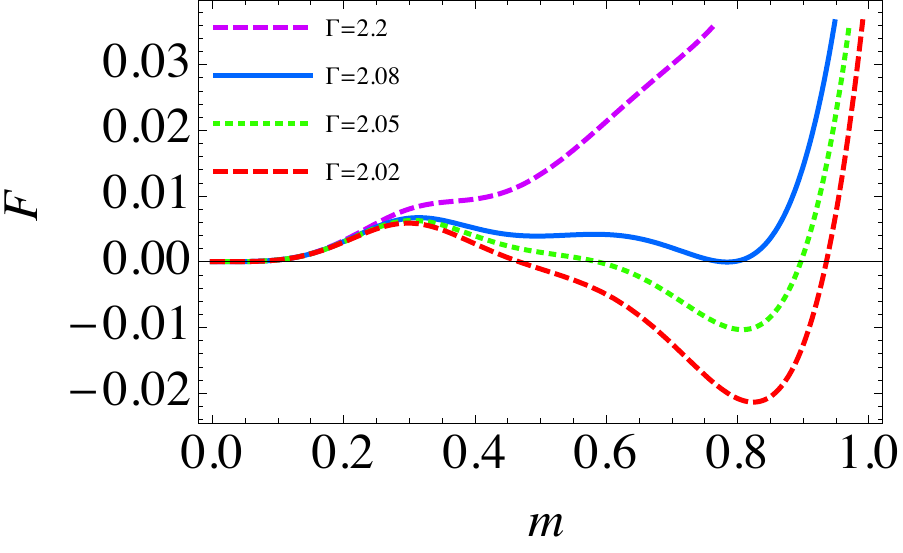}     \label{p4T01g07all} } &
 \subfigure[]{  \includegraphics[scale=0.6]{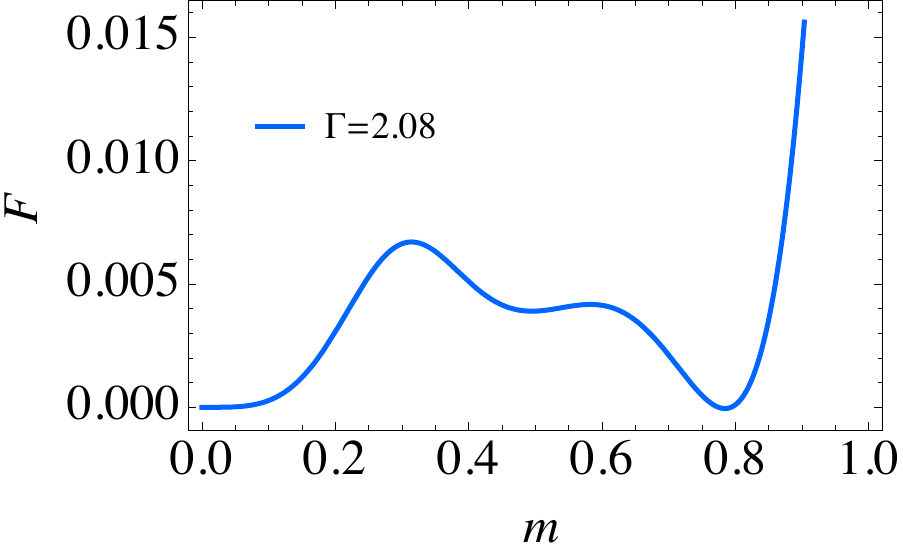}    \label{p4T01g07G208}} 
&   
 \subfigure[]{ \includegraphics[scale=0.6]{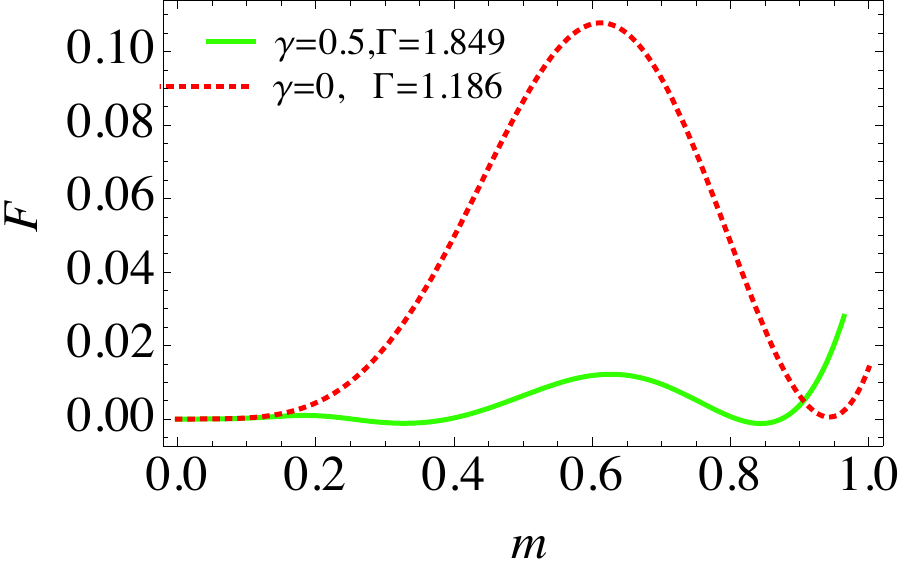} \label{p4T0025gcomp}}   
\end{tabular}     
\caption{ 
Free energy as a function of the order parameter $m$ for $p=4$.
$T=0.025$ in (a)-(f) with $\gamma=0.5$ in (a)-(c), and $\gamma=0.8$ in (d)-(f). (a) Four different $\Gamma$ values for $\gamma=0.5$, with two separated for clarity in (b) and (c);
(b) $\Gamma=1.91$,  (c)  $\Gamma=1.846$. (d) Four different $\Gamma$ values for $\gamma=0.8$, with two separated for clarity in (e) and (f); (e) $\Gamma=2.95$, (f) $\Gamma=2.22$.  In (g) and (h) $T=0.1$, $\gamma=0.7$. (g) Four different $\Gamma$ values, with $\Gamma=2.08$ separated for clarity in (h). (i) Again $T=0.025$. The green solid line is
  for $\gamma=0.5$ [Fig.~\ref{p4T0025g05G1846}] and the red dotted line is for $\gamma=0$. The potential barrier for $\gamma=0$
  is much larger than that of $\gamma=0.5$. 
See text for analysis.
}
\label{p4T0025g05}
\end{figure*}

Finally, we consider case (iii), with $(T,\gamma)=(0.1,0.7)$.
The free energy for various values of $\Gamma$ is shown in Fig.~\ref{p4T01g07all}.
As $\Gamma$ decreases, there is a phase transition from $m=0$ to $m_{\text{large}}$. 
There is a bump in the free energy around $m=m_{\text{small}}\simeq0.45$ Fig.~\ref{p4T01g07G208}.
However, the local minimum at $\ml$ reaches the value $F(0)$ first and the phase transition between $m=0$
and $\ml$ takes place before any other transition between $m=0$ and $\ms$ happens.
Therefore, this is the only first order phase transition.

\section{Perturbative analysis in the presence of a penalty transverse field}
\label{app:pert-th}

We study the phase transition using perturbation theory. Specifically, we compute the energy gap between the ground state and the first excited state of the effective single-body Hamiltonian. 

\bes
\begin{align}
\label{effective H transverse}
H_{\mathrm{eff}}&=H_0+V\ , \\
H_0&=\sum_{c=1}^{C}H_c,~~H_c=-pm^{p-1}\sigma^{z}_{c}-\gamma\sigma^{z}_{0}\sigma^{z}_{c}-\Gamma \sigma^{x}_{c}\ ,  \\
V&=-\epsilon \Gamma \sigma^{x}_{0}\ .
\end{align}
\ees

The eigenvalues of $H_c$ are $\pm E^\pm$, where 
\beq
E^\pm = \sqrt{(pm^{p-1}\pm\gamma )^2+\Gamma^2}\ ,
\label{eq:Epm}
\eeq 
and the superscript corresponds to the eigenvalues ($\pm 1$) of $\sigma^{z}_{0}$. We denote the eigenstates of $H_c$ by $\ket{c^+_\pm 0}$ with respective eigenvalues $\pm E^+$, and $\ket{c^-_\pm 1}$, with respective eigenvalues $\pm E^-$. The second entry ($0$ or $1$) refers to the eigenstates of $\sigma^z_0$.
The eigenstates of $H_0$ are, correspondingly,  $\ket{1^+_\pm 2^+_\pm \cdots C^+_\pm 0}$ and $\ket{1^-_\pm 2^-_\pm \cdots C^-_\pm 1}$, where now the last entry refers to the eigenstates of $\sigma^z_0$.
The ground state of $H_0$ is,  for $m > 0$,
\begin{equation}
|\Psi_{\rm g}\rangle = \ket{1_-^+ 2_-^+ 3_-^+ \cdots C_-^+0}
\end{equation}
with
\begin{equation}
H_0 |\Psi_{\rm g}\rangle =-CE^+ |\Psi_{\rm g}\rangle. 
\end{equation}
This ground state $|\Psi_{\rm g}\rangle $ is non-degenerate in $m > 0$.
We treat the penalty transverse field as a perturbation and compute 
the energy change.
The first-order perturbation vanishes.  The second order contribution
comes from a flip of $\sigma^z_0$ from $+1$ to $-1$ and then back to $+1$
with all other states kept intact. Then the perturbative correction to the ground state energy is, for $m>0$
\begin{align}
\label{deltaE}
\Delta E&=\frac{\epsilon^2 \Gamma^2}{C(E^+ -E^-)} \ ,
\end{align}
where $E^\pm$ is given in Eq.~\eqref{eq:Epm}.
This $\Delta E$ is to be added to Eq.~\eqref{zero T free energy}. 
In this perturbative regime, for a given $\gamma,p,$ and $\Gamma$, $\Delta E$ is a monotonically decreasing function of $m$. 
Therefore the free energy at smaller $m$ will be lifted more than that at larger $m$ by the presence of the transverse field on the penalty qubit.  This is consistent with our observation that for a sufficiently large $\epsilon$, the transverse field on the penalty qubit removes the additional local minimum $m_{\text{small}}$.

In the case of $m=0$, the ground state is doubly degenerate:
\bes
\begin{align}
|\Psi_{\rm g,1}\rangle &= \ket{1_-^+ 2_-^+ 3_-^+ \cdots C_-^+0} \ , \\
|\Psi_{\rm g,2}\rangle &= \ket{1_-^- 2_-^- 3_-^- \cdots C_-^-1}
\end{align}
\ees
We can then perform degenerate perturbation theory as long as $\epsilon \ll \Gamma$  to obtain:
\beq
\Delta E_{m=0} = \epsilon 2 \Gamma\,.
\eeq
We can argue that the value above is a minimum, as a function of $m$, for the contribution of the transverse field on the penalty qubits to the free energy. To see this, notice that first, that the free energy is symmetric under $m \leftrightarrow -m$ even in the presence of the penalty transverse field, so $m=0$ should be a local extremum. Second, $m=0$ should be a local minimum, since the penalty transverse field aligns the penalty qubits along the $x$ direction, and this favors the problems qubits to point in the $x$ direction, which is a $m=0$ state. Our analysis implies the existence of an intermediate value of $m$  where the free energy is maximally lifted. An important consequence of this is that the penalty transverse field increases the potential barrier between $m=0$ and $m_{\text{large}}$, and may remove the additional local minimum $m_{\text{small}}$.



\bibliography{refs}

\end{document}